\documentclass[printer]{aa}
\usepackage{graphicx}
\usepackage{hyperref}
\usepackage{amsmath}

\usepackage{natbib}
\bibpunct{(}{)}{;}{a}{}{,} 

\usepackage{txfonts}
\usepackage{longtable}
\usepackage{listings}
\usepackage{color}
\usepackage{textcomp}

\usepackage{subfig}

\begin{document}

\title{ Expanding associations in the Vela-Puppis region }

   \subtitle{3D structure and kinematics of the young population \thanks{The list of members of the young populations established in this study is only available in electronic form
at the CDS via anonymous ftp to cdsarc.u-strasbg.fr (130.79.128.5)
or via http://cdsweb.u-strasbg.fr/cgi-bin/qcat?J/A+A/}   }

\author{
T. Cantat-Gaudin\inst{\ref{IEECUB}}
\and
C. Jordi\inst{\ref{IEECUB}}
\and
N. J. Wright\inst{\ref{Keele}}
\and
J. J. Armstrong\inst{\ref{Keele}}
\and
A. Vallenari\inst{\ref{OAPD}}
\and
L. Balaguer-N{\'u}{\~n}ez\inst{\ref{IEECUB}}
\and
P. Ramos\inst{\ref{IEECUB}}
\and
D. Bossini\inst{\ref{OAPD}}
\and
P. Padoan\inst{\ref{IEECUB}}
\and
V. M. Pelkonen\inst{\ref{IEECUB}}
\and
M. Mapelli\inst{\ref{OAPD},\ref{IAT}}
\and
R. D. Jeffries\inst{\ref{Keele}}
}

\institute{
Institut de Ci\`encies del Cosmos, Universitat de Barcelona (IEEC-UB), Mart\'i i Franqu\`es 1, E-08028 Barcelona, Spain\label{IEECUB}
\and
Astrophysics Group, Keele University, Keele, ST5 5BG, UK\label{Keele}
\and
INAF, Osservatorio Astronomico di Padova, vicolo Osservatorio 5, 35122 Padova, Italy\label{OAPD}
\and
Institut f\"ur  Astro- und Teilchenphysik, Universit\"at Innsbruck, Technikerstrasse 25/8, A--6020, Innsbruck, Austria \label{IAT}
}

\date{Received date / Accepted date }

\abstract
{The Vela-Puppis region is known to host the Vela~OB2 association as well as several young clusters featuring OB and pre-main-sequence stars. Several spatial and kinematic subgroups have been identified in recent years.}
{By grouping stars based on their positions and velocity, we can address the question of the dynamical history of the region and the mechanisms that drove stellar formation. The \textit{Gaia}~DR2 astrometry and photometry enables us to characterise the 3D spatial and 3D kinematic distribution of young stars and to estimate the ages of the identified components.}
{We used an unsupervised classification method to group stars based on their proper motions and parallax. We studied the expansion rates of the different identified groups based on 3D velocities and on corrected tangential velocities. We used theoretical isochrones to estimate ages.}
{The young stars can be separated into seven main groups of different ages and kinematical distribution. All groups are found to be expanding, although the expansion is mostly not isotropic. }
{The size of the region, the age substructure, and the anisotropic expansion rates are compatible with a prolonged period of star formation in a turbulent molecular cloud. The current kinematics of the stars cannot be explained by internal processes alone (such as gas expulsion).}


\maketitle{}



\section{Introduction}
It has been known for a long time that O and B stars are not distributed randomly on the sky \citep{Eddington1910,Kapteyn14,Rasmuson21} and that they form sparse groups of co-moving stars. The term ``association'' was introduced by \citet{Ambartsumian47}. Although OB associations are too sparsely populated to be gravitationally bound \citep{Bok34,Mineur39}, their internal velocity dispersion is small enough that they can be identified as overdensities in velocity space. While the historical name OB association is widely used in the literature, these groupings of young stars are known to include lower-mass stars and pre-main-sequence stars, following a continuous mass function \citep{Briceno07}.

Significant progress in the description and characterisation of OB associations was made when the Hipparcos mission opened a new era in the study of the solar neighbourhood \citep[e.g.][]{Brown97,Comeron98,deZeeuw99,Brown99,Hoogerwerf99,Kaltcheva00,Elias06,Caballero08,Bouy15} by providing parallaxes and precise proper motions for $\sim$100,000 stars, most of them brighter than $G\sim12$.
More recently, the unprecedented quality of the astrometry provided by the second \textit{Gaia} data release \citep[\textit{Gaia}~DR2,][]{GDR2content} has enabled detailed studies of the spatial and kinematic substructure of stellar associations within several hundreds of parsecs of the Sun \citep[e.g.][]{Kuhn18,Kounkel18,Damiani18,Karnath18,Zari18,Kos18orion}.

The Vela~OB2 association was originally reported by \citet{Kapteyn14}. The first description of its kinematics comes from \citet{deZeeuw99}, who noted that the space motion of the association is not clearly separate from the surrounding stars and the nearby (slightly older) open clusters NGC~2547 and Trumpler~10.
\citet{Pozzo00} identified through X-ray observations a group of pre-main-sequence stars (PMS) that are located around the massive binary system $\gamma^2$~Vel, which lies at a distance of 350 to 400\,pc from us. This group of young stars \citep[$\sim$10\,Myr,][]{Jeffries09}, sometimes referred to as the Gamma Velorum cluster \citep[or Pozzo~1 in][]{Dias02}, has been the object of many recent publications. Making use of spectroscopic observations from the \textit{Gaia}-ESO Survey \citep[][]{Gilmore12,Randich13}, \citet{Jeffries14} and \citet{Sacco15} have shown that two kinematic groups exist in the region. In a subsequent study making use of N-body simulations, \citet{Mapelli15} showed that the spatial and velocity distributions of these stars correspond to a supervirial spatially expanding structure. 

Several further studies have been published since the release of the first \textit{Gaia} catalogue \citep[][]{Gaia16prusti,GaiaDR1}. Using \textit{Gaia}~DR1 data astrometry, \citet{Damiani17} noted that the proper motions of the stars located in a few degrees around $\gamma^2$~Vel are mainly distributed in two groups of different age, with the significantly older one including NGC~2547.
\citet{Franciosini18} have shown that the kinematic populations identified from the \textit{Gaia}-ESO Survey radial velocities also present distinct proper motions and parallaxes, and \citet{Beccari18} noted that the region might host not two but at least four non-coeval kinematic groups, with a fragmented spatial distribution.
\citet{Armstrong18} studied the distribution of the stars coeval with the Gamma Velorum cluster, and compared it with the Vela~OB2 stars of \citet{deZeeuw99}, noting that the OB stars are not all located in high-density regions of young PMS stars.
Finally, \citet{CantatGaudin19vela} identified 11 distinct groups in 3D positional and velocity space, spread over more than a hundred parsecs, and confirmed that their overall structure is expanding. The centre of this structure appears devoid of stars, and its projection on the sky matches the large bubble known as the IRAS Vela Shell \citep[IVS,][]{Sahu92phdt}. The authors suggested that feedback from a slightly older stellar population might have driven the formation of both the Vela~OB2 stars and the IVS.

In this paper we make use of \textit{Gaia}~DR2 to study the 3D spatial and 3D velocity distribution of young stars over a large scale in the direction of the Vela and Puppis constellations.
This paper is organised as follows: in Sect.~\ref{sec:data} we describe our data selection, in Sect.\ref{sec:kinematics} we characterise the observed radial and tangential velocity distributions, in Sect.~\ref{sec:ages} we derive ages for the different observed populations, and in Sect.~\ref{sec:groups} we provide individual comments on the different populations. Sect.~\ref{sec:discussion} contains a discussion, and we conlcude in Sect.~\ref{sec:conclusion} with a summary of this study.

\section{Data} \label{sec:data}

\begin{figure*}[ht]
\begin{center} \resizebox{\hsize}{!}{\includegraphics[scale=0.5]{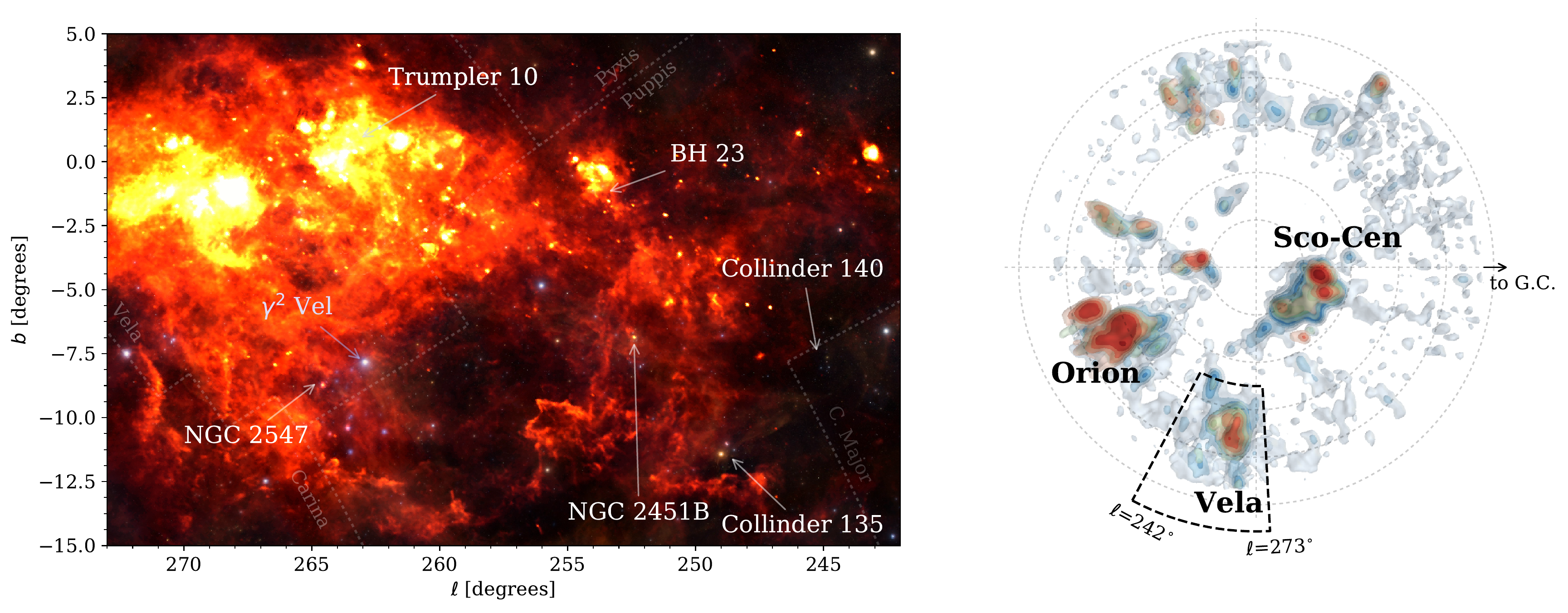}} \caption{\label{fig:zari} Left: Composite of an IRIS 60\,$\mu$m image \citep[][]{Miville05} and DSS optical image of the investigated region. Right: Distribution of PMS stars projected on the Galactic plane from \citet{Zari18}. The radii of the concentric circles centred on the Sun increase from 100 to 500\,pc. } \end{center}
\end{figure*}

We queried the \textit{Gaia}-DR2 archive\footnote{\url{https://gea.esac.esa.int/archive/}} for stars brighter than $G$=18 in a large field ($31^{\circ}\times20^{\circ}$) and in the parallax range $\varpi\in$[1.8,4]\,mas (roughly corresponding to distances from 250 to 550 pc). This spatial selection includes the known clusters Trumpler~10, NGC~2451B, NGC~2547, Collinder~135, Collinder~140, UBC~7, and BH~23, as well as the extended Vela~OB2 complex. The investigated region is shown in Fig.~\ref{fig:zari}.

We followed Eqs. (1) (restriction on the astrometric unit-weight error, UWE) and (2) (on the photometric excess flux) from \citet{Arenou18} in order to filter out the stars with poor astrometric solutions and poor photometry. 
This selection is slightly more permissive than the filter introduced by \citet{Lindegren18technicalnote}, who recommended to discard all sources with a re-normalised unit weight (RUWE) larger than 1.4. In the end (after applying the selections described in Sect.~\ref{sec:photometricselection} and Sect.~\ref{sec:upmask}), we find that about 4\% of our final sample has RUWE>1.4, and 1\% has RUWE>2, while using the RUWE filter would introduce no additional sources. The stars with RUWE>1.4 exhibit the same patterns in photometric, spatial, and velocity space as the rest of the sample (although often with larger astrometric uncertainties than stars at the same magnitude), and the results presented in this paper do not depend on the choice of filtering. 

Throughout this paper we estimate distances to individual stars by inverting their parallax. Although this simple estimation is not recommended for the vast majority of the sources in the \textit{Gaia}~DR2 catalogue \citep[see e.g.][]{BailerJones15,Luri18}, the stars we studied have typical fractional parallax uncertainties of 2\% (and are all under 11\%), which introduces only a negligible bias. All uncertainties on quantities derived from distances (such as absolute magnitude $M_G$ or tangential velocity $v$\footnote{We recall that $v\simeq4.74\mu/\varpi$, expressed in km\,s$^{-1}$ if proper motion $\mu$ in mas\,yr$^{-1}$ and parallax $\varpi$ in mas.}) are estimated by propagating parallax uncertainties (using the covariance matrix between the astrometric measurements $\mu_{\alpha*}$, $\mu_{\delta}$, and $\varpi$).

We note $M_G = G+5 \log_{10}\left( \frac{\varpi}{1000} \right) + 5$ (if $\varpi$ expressed in mas) the observed magnitude $G$ corrected for distance, but not for extinction.
We added a fixed correction of +0.029\,mas to all \textit{Gaia}~DR2 parallaxes. This value corresponds to the parallax zero-point offset reported by \citet{Lindegren18}, and roughly corresponds to a 2\,pc offset at a distance of 250\,pc, or a 10\,pc offset at a distance of 600\,pc. We recall that because of the uncertainty on the true value of this zero-point \citep[and its possible dependence on colour and magnitude,][]{Arenou18} and local systematics \citep{Lindegren18,Arenou18,Vasiliev18}, the uncertainty on the true absolute distance of a source is always larger than the uncertainty on its inverse parallax.

\subsection{Photometric selection} \label{sec:photometricselection}
We empirically devised a broad photometric selection that is intended to retain the younger stars while rejecting low-mass main-sequence stars and evolved stars where possible. The filter was built using the members of Trumpler~10, NGC~2451B, NGC~2547, Collinder~135, Collinder~140, and the Gamma Velorum cluster (Pozzo~1) as reference stars. The cluster member list for these objects was taken from \citet{CantatGaudin18gdr2}, who performed a selection based on astrometric criteria only.

\begin{figure}[ht]
\begin{center} \includegraphics[scale=0.55]{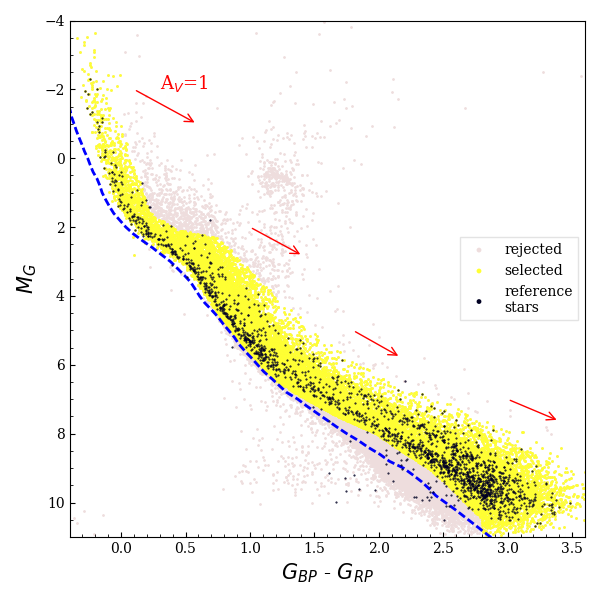} \caption{\label{fig:photoselection} Photometric selection of young stars. The reference stars are the members of the young clusters Trumpler~10, NGC~2451B, NGC~2547, Collinder~135, Collinder~140, and the Gamma Velorum cluster (Pozzo~1). The dashed line corresponds to the zero-age main sequence for a PARSEC isochrone of metallicity Z=0.019 and $A_V$=0.15.} \end{center}
\end{figure}

We defined the blue edge of the filter as the 10th percentile of the $G_{BP}-G_{RP}$ colour of the reference stars in sliding windows of 1\,mag in $M_G$, offset by -0.1\,mag. The red edge of the filter was arbitrarily chosen to discard evolved stars.
Figure~\ref{fig:photoselection} shows the location of the retained and discarded stars in a HR diagram. The filter discards about half the stars fainter than $M_G\sim7$, and a third of the total sample.

\begin{table*}[h]
\begin{center}
\caption{ \label{tab:oc_lit} Cluster parameters from the literature}
        \begin{tabular}{ l  c  c  c  c  c c }
        \hline
        \hline
cluster & $\ell$       & $b$ & distance         & RV   & $E(B-V)$               & $\log t$\\
        & [$^{\circ}$] & [$^{\circ}$]& [pc]     & [km\,s$^{-1}$] & mag          &         \\
        \hline
        BH~23           & 253.97 & -1.03  & 438         & 17.4$^a$        & 0.05$^a$     & 7.14$^a$ \\
                        &               &   &   & 18$^b$          & 0.06$^b$         & 8.4$^b$ \\
                        &               &   &   & 22.83$^c$  & -        & - \\
        \hline
        Collinder~135   & 248.99 & -11.20 & 302         & 15.4$^a$  & 0.042$^a$         & 7.6$^a$ \\
                        &       &        &      & 15.85$^b$  & 0.032$^b$         & 7.41$^b$ \\
                        &       &        &      & 16.03$^c$  & -                & - \\
        \hline
        Collinder~140   & 244.94 & -7.82   & 381        & 22.9$^a$  & 0.052$^a$         & 7.7$^a$ \\
                        &               &   &   & 19.34$^b$  & 0.03$^b$         & 7.55$^b$ \\
                        &               &   &   & 19.65$^c$  & 0.034$^e$         & 7.47$^e$ \\

        \hline
        NGC~2451B       & 252.31 & -6.86  &  364        & 13.7$^a$   & 0.167$^a$         & 8.23$^a$ \\
                        &               &   &   & 14$^b$     & 0.055$^b$         & 7.65$^b$ \\
                        &        &   &          & 16.78$^c$  & 0.067$^e$         & 7.59$^e$\\

        \hline
        NGC~2547        & 264.445 & -8.60  & 387        & 14.8$^a$  &  0.04$^a$         & 7.89$^a$ \\
                        &        &   &          & 15.65$^b$  & 0.186$^b$         & 7.59$^b$ \\
                        &        &   &          & 14.07$^c$  & 0.04$^e$         & 7.43$^e$ \\

        \hline
        Trumpler~10     & 262.87 & 0.58  &  432 & 15$^a$     & 0.029$^a$         & 7.38$^a$ \\
                        &       &       &       & 32.17$^b$  & 0.034$^b$         & 7.54$^b$ \\
                        &        &       &      & 22.43$^c$  & 0.056$^e$         & 7.65$^e$ \\

        \hline
        Gamma Vel.      & 262.798 &-7.691  & 347        & 16.84$^d$  & 0.057$^e$         & 7.18$^e$\\

        \hline
        \hline
        \end{tabular}
        
\tablefoot{ $\ell$, $b$, and distance from \citet{CantatGaudin18gdr2}. Other references: $^a$\citet{Kharchenko13} ; $^b$\citet{Dias02} ; $^c$\citet{Soubiran18} ; $^d$\citet{CantatGaudin19vela}; $^e$\citet{Bossini19}. } 
\end{center}
\end{table*}

\subsection{Astrometric cleaning of the field stars using UPMASK} \label{sec:upmask}
The \textit{Gaia}~DR2 astrometry allows us to identify groups of stars not solely as overdensities in positional space, but as sharing common proper motions ($\mu_{\alpha*}$,$\mu_{\delta}$) and parallaxes ($\varpi$) as well. In this study we applied the unsupervised classification scheme of UPMASK \citep{KroneMartins14}. This approach does not rely on assumptions on the structure of a group of star in positional or in astrometric space, and only requires that stars with similar properties in astrometric space must occupy less space on the sky than a random uniform distribution (we call this step the veto). In our implementation \citep[as in][]{CantatGaudin18tgas}, we used $k$-means clustering to define small groups of stars in astrometric space, and the total length of a minimum spanning tree (MST) to compare the spatial distribution of each group with a uniform distribution.
We consider that a group of stars is clustered if the total branch length of its MST is shorter than the mean MST of a random distribution by at least 1$\sigma$.

We repeated the procedure 100 times, each time redrawing a new value for the proper motion and parallax of every star based on the full covariance matrix of its ($\mu_{\alpha*}$,$\mu_{\delta}$,$\varpi$) uncertainties. In a frequentist approach, the final score (how many times a star was part of a group that passed the veto) can be interpreted as a membership probability ranging from 0 to 100\%.

The advantage of this approach is that we do not require groups to be spatially dense, but only that they occupy a smaller area than the whole field of view. This allowed us to retain elongated structures if they are made up of stars with similar proper motions and parallaxes. The approach is independent of the number of clusters present, as it only discards non-clustered stars. It is also entirely unsupervised and does not rely on a threshold density or a maximum distance between points in the parameter space \citep[as in e.g. DBSCAN,][]{Ester96dbscan}, which allowed us to find clusters of various densities, and to keep the low-density tails of the identified groups.

\subsection{Splitting the sample into distinct populations} \label{sec:split}

\begin{figure*}[ht]
\begin{center} \includegraphics[scale=0.8]{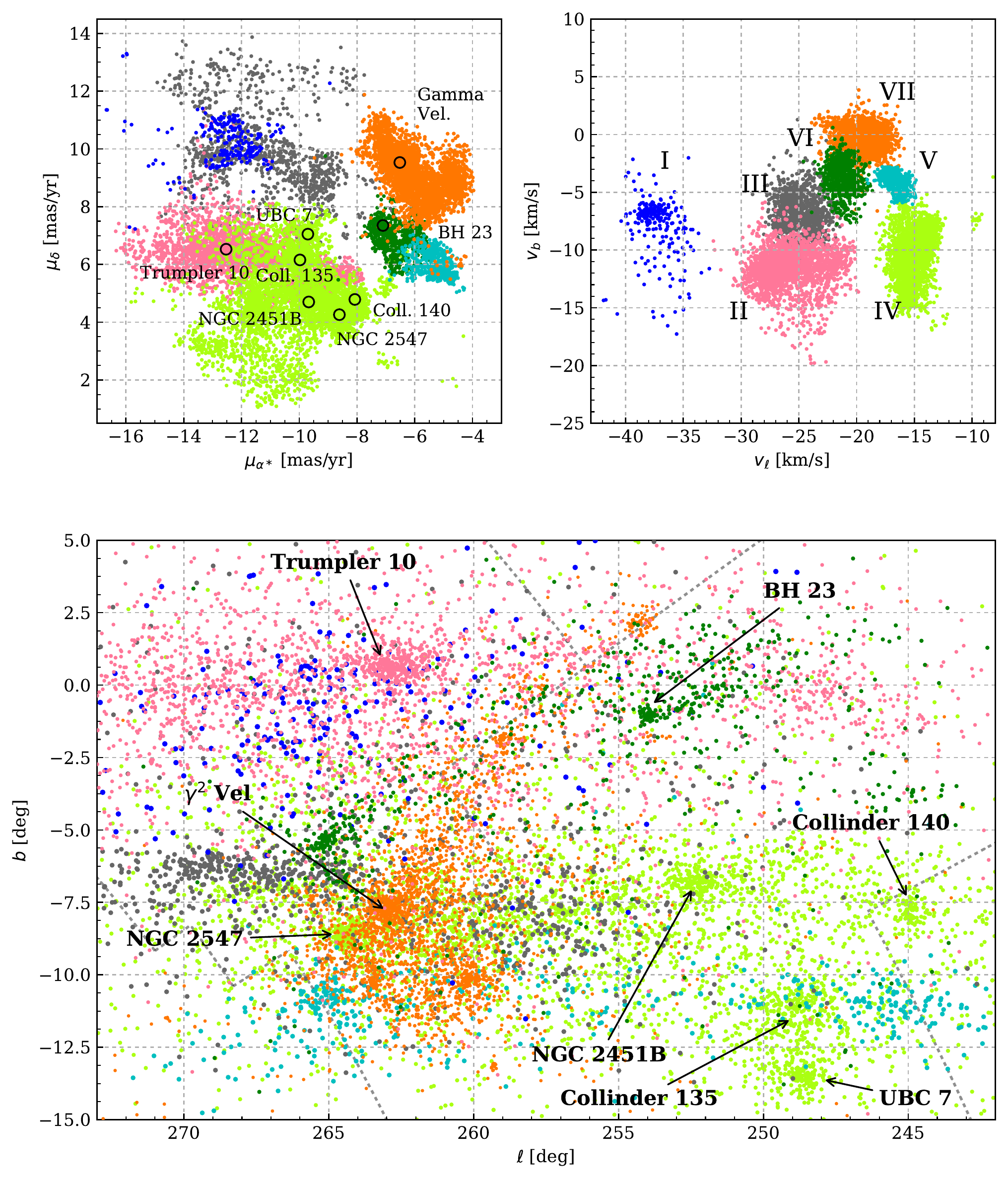} \caption{\label{fig:propermotionspace} Top left: Proper motions of the stars with UPMASK clustering scores larger than $50\%$. Top right: Tangential velocities (in Galactic coordinates) of the same stars. The colours correspond to the seven populations defined from their tangential velocities (see text). Bottom: Spatial distribution of the same stars.} \end{center}
\end{figure*}

We verified that the stars with membership scores $<50\%$ were uniformly distributed in the field of view and showed no obvious structure in proper motion space. In the rest of this study we only worked with the sample of stars with scores $>50\%$, whose proper motions ($\mu_{\alpha*}$,$\mu_{\delta}$), tangential velocities ($v_{\ell}$,$v_{b}$), and positions ($\ell$,$b$) are shown in Fig.~\ref{fig:propermotionspace}. 

The distribution in proper motion space is partly shaped by the fragmented spatial distribution of the stars, and their distribution in tangential velocity space (accounting for the individual parallax of each star) appears more compact, as shown in Fig.~\ref{fig:propermotionspace}. From Monte Carlo propagation of the \textit{Gaia}~DR2 uncertainty on the proper motions and parallaxes, we find the typical uncertainty in tangential velocity to be $\sim0.3$\,km\,s$^{-1}$. 

We identified seven main groups in ($v_{\ell}$,$v_{b}$) space, showing very little overlap. We labelled these groups from population I to population VII (in order of decreasing age, as later estimated in Sect.~\ref{sec:ages}). Their distributions on the sky ($\ell$,$b$) and on the plane of the Milky Way ($X$,$Y$) are shown in Figs.~\ref{fig:map_Na} to \ref{fig:map_V}.

We note that our sample includes a few upper main-sequence stars from the nearby \citep[$\sim$278\,pc,][]{CantatGaudin18gdr2} older cluster Alessi~3 \citep[$\log t$ of 8.7 and 8.87 in][respectively]{Dias02,Kharchenko13}, which is located less than half a degree outside of the southern boundary of the investigated region. These stars can easily be discarded based on their tangential velocity and compact location on the sky.

We tried several approaches in order to provide the final division of the sample into seven populations. We found that the two methods that provide the most satisfying decomposition are i) performing hard cuts in ($v_{\ell}$,$v_{b}$) and ($\mu_{\alpha*}$,$\mu_{\delta}$) space or ii) agglomerative clustering \citep{Mullner11}. Gaussian mixture models or k-means clustering performed poorly on this dataset, overfitting the observed structure or grouping together stars with too different velocities. The division we adopted for the rest of this study was obtained from agglomerative clustering. We note that most stars were assigned the same population regardless of the method used, and that the choice of splitting method only concerns a small number of stars. We verified that the results and conclusions of this paper are identical whether supervised agglomerative clustering or purely manual splitting was performed. In particular, the velocities of the five clusters known as Collinder~135, Collinder~140, NGC~2451B, NGC~2547, and UBC~7 are similar enough for the clusters to always be classified as subgroups of the same main kinematic population.

\section{Linear expansion rates} \label{sec:kinematics}
In this section we characterise the kinematic structure of the seven populations by establishing whether they exhibit a spatial velocity gradient. A positive correlation between the position along a given axis and the velocity along this axis is interpreted as evidence for expansion. The approach we follow is similar to that applied by \citet{Wright18} to the Scorpius-Centaurus association.

\subsection{From 3D velocities} \label{sec:rv} 
\begin{figure*}[ht]
\begin{center} \includegraphics[scale=0.7]{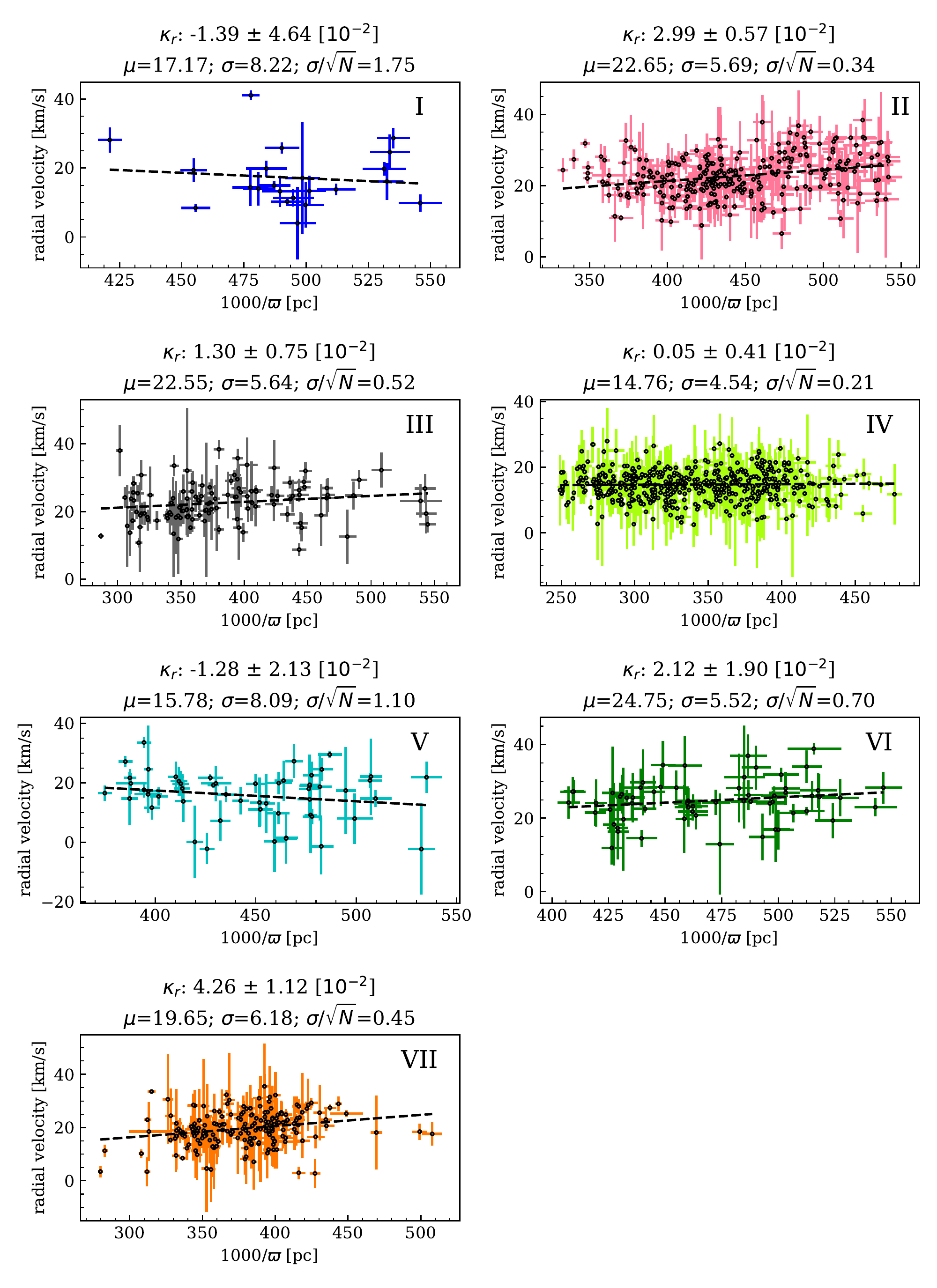} \caption{\label{fig:radial_expansion} \textit{Gaia}~DR2 radial velocity as a function of distance for the stars in each population identified in this study. $\kappa_r$: slope of the radial velocity gradient with distance ($\times 10^{-2}$\,km\,s$^{-1}$\,pc$^{-1}$). $\mu$: mean radial velocity (km\,s$^{-1}$). $\sigma$: standard deviation (km\,s$^{-1}$). $\sigma$/$\sqrt{N}$: statistical uncertainty (km\,s$^{-1}$).} \end{center}
\end{figure*}

The most direct way to study the internal kinematics of a group of stars is to independently determine a 3D velocity vector for each member star. In practice, only 10\% of the stars in our sample have \textit{Gaia}~DR2 radial velocities, with a typical uncertainty of 3.3\,km\,s$^{-1}$. Figure~\ref{fig:radial_expansion} shows the observed \textit{Gaia}~DR2 radial velocities as a function of line-of-sight distance for the seven populations. We estimated the uncertainty on the slope by Monte-Carlo redrawings of parallaxes and radial velocities of the Gaussian nominal uncertainties. The observed gradient is only significantly positive (by more than 2$\sigma$) for populations II and VII because of the relatively large nominal uncertainties on the radial velocities.

\begin{figure}[ht]
\begin{center} \includegraphics[scale=0.8]{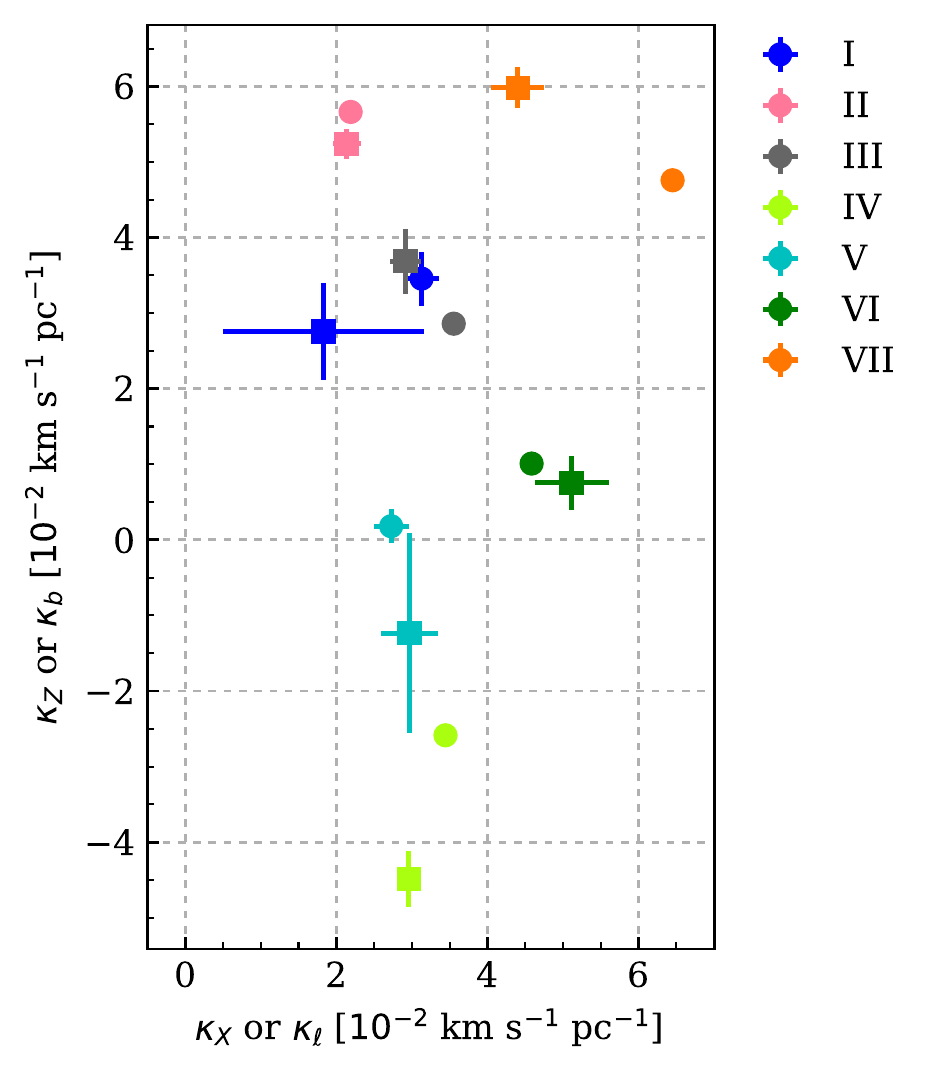} \caption{\label{fig:lineargradients} Observed linear expansion rates (as listed in Table~\ref{table:slopes}). Squares: ($\kappa_X$,$\kappa_Z$) from 3D velocities. Dots: ($\kappa_{\ell}$,$\kappa_b$) from corrected tangential velocities (see Sect.~\ref{sec:vtan}). The colour-coding is the same as in Figs.~\ref{fig:propermotionspace} and \ref{fig:radial_expansion}. } \end{center}
\end{figure}

We computed the position ($X$,$Y$,$Z$) in Galactic Cartesian coordinates for each member of each population, and its velocity vector (v$_X$,v$_Y$,v$_Z$) from its observed proper motions and radial velocity. For each component we discarded the stars whose velocity was discrepant from the sample mean by more than 3$\sigma$, then fit a linear relationship between $X$ and v$_X$ (and similarly for the $Y$ and $Z$ axes). The slopes ($\kappa_X$, $\kappa_Y$, and $\kappa_Z$) we obtained are listed in Table~\ref{table:slopes} and are shown in Fig.~\ref{fig:lineargradients} (for the $X$ and $Z$ axes only). Because the investigated region is located near $\ell$=270$^{\circ}$, the velocity component $v_Y$ is roughly equivalent to the radial velocity and inherits its large uncertainty, while $v_X$ and $v_Z$ are better constrained.

We observe that all seven populations exhibit a significant positive correlation between $X$ and $v_X$ ($\kappa_X$>0). Most of them also exhibit signs of expansion along the $Z$ axis (perpendicular to the Galactic plane), with the exception of population V, which exhibits no significant correlation, and population IV, whose velocity gradient $\kappa_Y$ appears negative. Rather than a sign of contraction, we argue that a simple linear expansion model is not appropriate to describe the internal kinematics of this group, which exhibits a complex substructure, and whose individual components would be better treated separately.
We further discuss this population in Sect.~\ref{sec:groups}. Overall, the only groups exhibiting an isotropic expansion ($\kappa_X \simeq \kappa_Z$) are populations I and III.

Following the formulation of \citet{Blaauw64} and \citet{Wright18}, the slopes $\kappa$ can also be expressed as an expansion age: $\tau = (\gamma \kappa)^{-1}$ (where $\gamma$=1.0227 is the conversion factor from km\,s$^{-1}$ to pc\,Myr$^{-1}$. Inverting this relation allows us to derive $\kappa$ from an estimated age. We find that the observed $\kappa_X$ (and $\kappa_\ell$) are slightly lower than expected from the ages derived in Sect.~\ref{sec:ages}, but that they generally follow the expected trend, with the youngest (populations VII and VI) exhibiting the highest expansion rates and populations I and II the slowest.

\subsection{From tangential velocities} \label{sec:vtan}
The precision of the \textit{Gaia}~DR2 astrometry allows us to derive tangential velocities ($v_{\ell}$,$v_{b}$) with typical uncertainties under 0.3\,km\,s$^{-1}$ for a large number of stars. Because most of these stars lack a measurement of their radial velocity, the interpretation of these tangential velocities is not straightforward.

The observed proper motions (and inferred tangential velocities) of the stars that move away from us suffer from a perspective effect that introduces a virtual contraction (for a group of stars with negative radial velocity, a virtual expansion). If the bulk radial velocity of the group of stars is known, the effects of virtual contraction and of true physical contraction can be distinguished. When no physical expansion or contraction is assumed, the mean distance can be inferred from the radial velocity with the so-called moving cluster method \citep[see e.g.][]{Jones71,deBruijne99}. Alternatively, a mean radial velocity can be derived for a non-expanding cluster of known distance \citep[e.g.][]{Dravins99}. We corrected the tangential velocities ($v_{\ell}$,$v_{b}$) following equation (A3) of \citet{Brown97}. The corrected tangential velocities ($v'_{\ell}$,$v'_{b}$) are

\begin{equation} \label{eq1}
\begin{split}
v'_{\ell} & = v_{\ell} - v_{rad} \cos b_c \sin( \ell_c - \ell ) , \\
 v'_{b} & = v_{b} - v_{rad} \left[  \sin b_c \cos b  - \sin b \cos b_c \cos \left( \ell_c - \ell  \right) \right]
\end{split}
,\end{equation}

\noindent where ($\ell_c$,$b_c$) is the location of the centre of the distribution of each population, and $v_{rad}$ is the mean radial velocity. 
To a good approximation ($\ell_c - \ell < 15^{\circ}$), virtual contraction introduces a linear expansion with angular distance from the distribution centre \citep[see e.g.][]{Blaauw52,Lesh68}.

After applying the correction for virtual expansion using the mean radial velocities derived in Sect.~\ref{sec:rv}, we find that all populations exhibit significant residual positive gradients of $v'_{\ell}$ with $\ell$ (and similarly of $v'_{b}$ with $b$). We only illustrate the apparent and perspective-corrected gradients for population II in Appendix~\ref{appendix:vtan}. We show the final velocity maps for all seven populations in Figs.~\ref{fig:map_Na} to \ref{fig:map_V}. 

For a meaningful comparison with the linear expansion slopes derived from 3D velocities in Sect.~\ref{sec:rv}, we report in Table~\ref{table:slopes} and show in Fig.~\ref{fig:lineargradients} the slopes $\kappa_\ell$ and $\kappa_b$ obtained after converting the angular distance from the distribution ($\ell_c - \ell$) into a physical distance. Because the centre of the region investigated in this paper is located close to ($\ell$,$b$)=(270$^{\circ}$,0$^{\circ}$), $\kappa_\ell$ and $\kappa_b$ roughly correspond to $\kappa_X$ and $\kappa_Z$ (respectively).
We note that in all cases the two methods yield very similar expansion rates, and in particular, the methods show that all groups expand in the $\ell$ direction. The two groups for which the two results differ most are populations IV and VII, which happen to be the two groups with the most spatial and kinematic substructure.

We attempted to characterise the expansion of the subgroups of populations III, V, and VI, and found that they follow the same expansion pattern as their overall population. The uncertainties on the slopes $\kappa$ obtained for subsets, however, are larger than for the whole population because we have fewer points for them and the spatial baseline that is available to constrain the velocity gradient is shorter.

        \begin{table*}[h!]
        \begin{center}
        \caption{ \label{table:slopes} Linear expansion rates}
                \begin{tabular}{ l  c  c  c |   c  c  c }
                \hline
                \hline
pop.    & $\kappa_X$ & $\kappa_Y$ & $\kappa_Z$  & $\kappa_\ell$ & $\kappa_r$ & $\kappa_b$ \\ 
                \hline
I       &  1.83 $\pm$ 1.33   & -2.13 $\pm$ 4.45   & 2.76 $\pm$ 0.64   & 3.13 $\pm$ 0.23   & -1.39 $\pm$ 4.64   & 3.46 $\pm$ 0.36 \\
II      &  2.14 $\pm$ 0.19   & 0.98 $\pm$ 0.57   & 5.24 $\pm$ 0.19   & 2.19 $\pm$ 0.06   & 2.99 $\pm$ 0.57   & 5.66 $\pm$ 0.09 \\
III     &  2.92 $\pm$ 0.21   & 0.35 $\pm$ 0.75   & 3.69 $\pm$ 0.43   & 3.55 $\pm$ 0.13   & 1.30 $\pm$ 0.75   & 2.86 $\pm$ 0.12 \\
IV      &  2.96 $\pm$ 0.14   & 1.60 $\pm$ 0.36   & -4.49 $\pm$ 0.37   & 3.44 $\pm$ 0.07   & 0.05 $\pm$ 0.41   & -2.59 $\pm$ 0.08 \\
V       &  2.97 $\pm$ 0.37   & 2.53 $\pm$ 1.75   & -1.23 $\pm$ 1.32   & 2.73 $\pm$ 0.23   & -1.28 $\pm$ 2.13   & 0.18 $\pm$ 0.22 \\
VI      &  5.11 $\pm$ 0.49   & 3.39 $\pm$ 1.68   & 0.75 $\pm$ 0.36   & 4.58 $\pm$ 0.15   & 2.12 $\pm$ 1.90   & 1.01 $\pm$ 0.16 \\
VII     &  4.40 $\pm$ 0.35   & 5.68 $\pm$ 1.10   & 5.98 $\pm$ 0.27   & 6.45 $\pm$ 0.14   & 4.26 $\pm$ 1.12   & 4.76 $\pm$ 0.11 \\
                \hline
                \hline
                \end{tabular}
        
        \tablefoot{ $\kappa$ in 10$^{-2}$\,km\,s$^{-1}$\,pc$^{-1}$.} 
        \end{center}
        \end{table*}

\section{Ages of the different components} \label{sec:ages}

\begin{figure*}[ht]
\begin{center} \resizebox{\hsize}{!}{\includegraphics[scale=0.5]{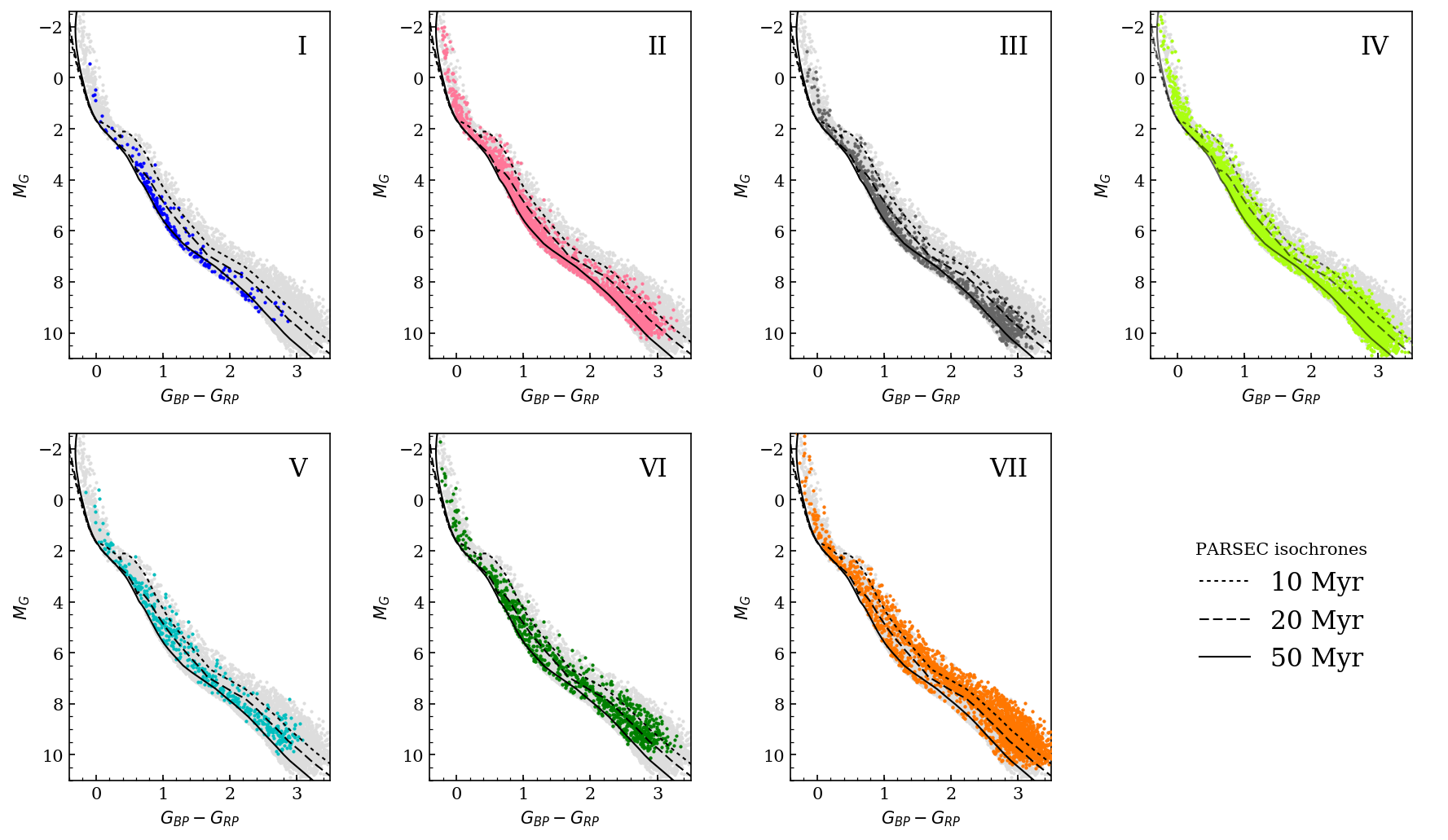}} \caption{\label{fig:cmd_populations} HR diagrams for the seven populations, along with three PARSEC isochrones computed for $A_V$=0.15 and $Z$=0.019. In all panels the light grey points show the entire sample.  } \end{center}
\end{figure*}

The Hertzsprung-Russel (HR) diagrams of the seven identified populations are shown in Fig.~\ref{fig:cmd_populations}. This figure also displays PARSEC (version 1.2S) isochrones \citep{Bressan12} of metallicity Z=0.019, computed for an absorption $A_V$=0.15 (compatible with the various estimates of absorption found in the literature for the known clusters in the region, see Table~\ref{tab:oc_lit}) with the \textit{Gaia}-DR2 passbands of \citet{Evans18}. 

The sequences of the different populations align with isochrones of different ages. We verified that sources fainter than $M_G$=5.5 and bluer than a $\log t$=7.5 isochrone in populations VI and VII are uniformly distributed across the investigated field of view. They therefore mostly correspond to contamination by main-sequence field stars and unresolved binaries. The remaining field star contamination in the older populations unfortunately overlaps too much with the PMS and cannot be easily identified on the basis of photometry.

We spatially divided the components of each population into several subgroups in order to estimate ages individually. For populations II, III, V, and VI, we manually split the population into two (eastern and western) subgroups. For population VI we selected the stars that lie within 1$^{\circ}$ of the centres of Collinder~135, Collinder~140, NGC~2451B, NGC~2547, and UBC~7 as their own subgroups. To split population VII (Vela~OB2), we selected the stars that lie within 1$^{\circ}$ of the groups A, C, D, G, and H defined in \citet{CantatGaudin19vela}. These divisions are marked on the maps of Appendix~\ref{appendix:maps}.

\begin{figure}[ht]
\begin{center} \resizebox{\hsize}{!}{\includegraphics[scale=0.5]{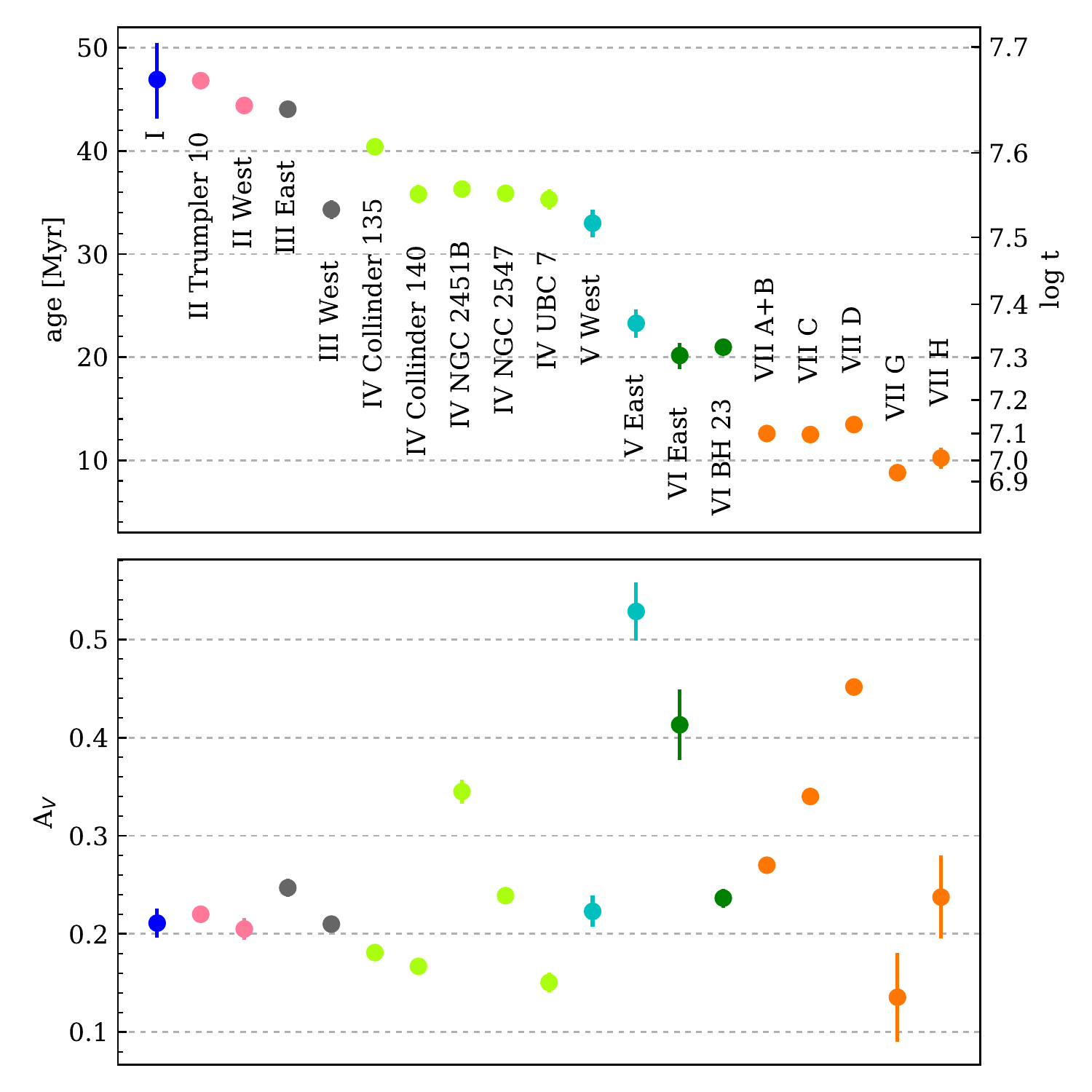}} \caption{\label{fig:ages} Ages (top) and extinction $A_V$ (bottom) of the different subgroups, estimated from isochrone fitting. The error bar represents the statistical uncertainty and is smaller than the marker size for most groups. } \end{center}
\end{figure}

We performed a maximum likelihood fitting of PARSEC isochrones through the observed $M_G$ and $G_{BP}-G_{RP}$ for each population using a metallicity of Z=0.019 and letting extinction and age be free parameters. The isochrones were reddened following the polynomial coefficients given in \citet{GDR2hrd}. The resulting best-fit values are shown in Fig.~\ref{fig:ages}. The procedure takes into account nominal photometric errors as well as the uncertainty on $M_G$ due to the parallax uncertainty. We remark that due to the millimagnitude quality of the photometry, the statistical uncertainty on the result is very small. We estimated the uncertainty on the best-fit age by making different samples of each subgroup, first based on the membership scores of Sect.~\ref{sec:upmask} and then by bootstrapping redrawings. In all cases, the uncertainties on the best-fit age are below 0.05 in $\log t$. This value does not reflect our actual uncertainty on the true age of the groups, which is dominated by the uncertainty on stellar evolution models and on the characterisation of the \textit{Gaia} passbands. We find that using PARSEC isochrones computed with the passbands of \citet{MaizApellaniz18phot} increases all estimated $\log t$ by $\sim$0.05 and the passbands of \citet{Weiler18} by almost 0.1.

Although the interpretation of colour-magnitude diagrams for PMS stars and the presence of unresolved binaries can imitate age spreads \citep[][]{Hillenbrand08,Baraffe09}, our simple fitting procedure highlights the fact that the different groups identified in this study are not coeval.
The age derived here for the youngest groups of population VII are in agreement with the values of 7.5$\pm$1\,Myr derived by \citet{Jeffries17} for the Vela~OB2 association from optical photometry.
The authors have shown, however, that lithium depletion patterns for these stars correspond to an older age (of 18 to 21\,Myr) and that photometric ages of very young stars are likely underestimated. In any case, an age of $\sim$20\,Myr would still make populatoin VII the youngest group in the sample investigated in this paper.

Because the reddening vector is mostly parallel to the pre-main sequence, extinction and age are not strongly degenerate. We remark, however, that the best-fit values of $A_V$ can be made lower when isochrones of higher metallicity are employed. This does not strongly affect the ages, in particular not the relative age distribution between the different subgroups. 

The eastern groups of populations V and VI (both located farther away than 450\,pc) exhibit a significantly higher extinction. This agrees with their position in the maps shown in Appendix~\ref{appendix:ag}.

\section{Comments on each population} \label{sec:groups}

\subsection*{Population I}
This is the least populated and appears to be the oldest of the seven identified groups. It is also one of the most distant groups and occupies the north-east quarter of the investigated region in the distance range 450-550\,pc (Fig.~\ref{fig:map_Na}). 

Because its position correlates with the region of higher dust emission that is visible in Fig.~\ref{fig:zari}, it is possible that this group is significantly affected by contamination from reddened zero-age main-sequence (ZAMS) stars. The location of this group in positional space and velocity space does correspond to an overdensity of blue stars (with $G_{BP}-G_{RP}$<0.3), however, and the upper main-sequence stars of population I do not appear to be significantly redder than the rest of the sample.

Population I presents a very low contrast against field stars in spatial and astrometric space. Collecting additional information (for instance, additional radial velocities) might allow identifying more sources as associated with population I.

\subsection*{Population II}
This group includes the known cluster Trumpler~10 and many stars that are mostly located in the northern half of the investigated region ($b>-5^{\circ}$) at distances larger than 350\,pc (Fig.~\ref{fig:map_Oa}). The distribution appears continuous, with a high overdensity around the location of Trumpler~10 ($\sim$420\,pc) and a secondary overdensity in the north-west region ($\sim$460\,pc). Although this group is very elongated along the $\ell$ direction, its expansion rate (as modelled with a linear gradient) appears to be higher in the $b$ direction (Fig.~\ref{fig:lineargradients}).

\subsection*{Population III}
This group is mainly distributed in the southern half of the region ($b<-5^{\circ}$) in two main subgroups (Fig.~\ref{fig:map_Nb}). The eastern subgroup is denser and elongated. The western, sparser subgroup appears to occupy the cavity inside the ring-like structure reported by \citet{CantatGaudin19vela}. Its projection also coincides with the structure known as the IRAS Vela Shell, and it could be the host population of the supernova that initiated the expansion of the shell.

\subsection*{Population IV}
This group hosts the most stars. It stretches across the whole southern half of the region (Fig.~\ref{fig:map_Ob}). It contains the known objects NGC~2547, NGC~2451B, Collinder~135, UBC~7, and Collinder~140. The dense clump that is visible in the top panel of Fig.~\ref{fig:map_Ob} about two degrees to the south of Collinder~135 is the object reported as UBC~7 by \citet{Castro18}, for which the authors suggested a common origin with Collinder~135.
The elongated structure that appears to bridge NGC~2547 to NGC~2451B is in fact located $\sim$100\,pc closer than these two clusters. Its proper motions match the group identified as Cluster~6 by \citet{Beccari18}.

Population IV is especially interesting because its five densest clumps correspond to known clusters. Several authors have searched for pairs or multiple systems of open clusters \citep{Piskunov06,delaFuenteMarcos09,Vazquez10,Conrad17} and proposed groupings based on statistical similarities between observed parameters. \citet{Soubiran18} have proposed that Collinder~140 and NGC~2451B are related objects. In this study we show that NGC~2547, NGC~2451B, Collinder~135, UBC~7, and Collinder~140 not only share a common space motion, but are still physically connected by a continuous distribution of stars. We also show in Fig.~\ref{fig:map_extended_dense_cmd} that the cluster stars and the extended population overlap perfectly in a colour-absolute magnitude diagram and therefore show no appreciable age difference.
The middle panel of Fig.~\ref{fig:map_Ob} shows that this population exhibits a significant level of kinematic substructure in addition to its radial expansion.

\subsection*{Population V}
This group stretches along the southern border of the region (Fig.~\ref{fig:map_VVb}) and shows two overdensities in the south-east and south-west corners. The eastern subgroup corresponds to group J from \citet{CantatGaudin19vela}. 

The western subgroup is older, sparser, and is not visible in the final density maps of \citet{CantatGaudin19vela} because the photometric selection and proper motion cuts applied in that study were designed to target the youngest known population (Vela~OB2), while the slightly younger eastern subgroup is also more reddened. The two subgroups have slightly different $v_b$ velocities. The age and $v_b$ of the western subgroup are closer to the age and velocity of population IV, while the age and velocity of the eastern group are more similar to those of the younger groups.

Population V exhibits no significant expansion in the $b$ (or $Z$ direction). It appears to expand only along the direction of $\ell$ (see gradients in Fig.~\ref{fig:lineargradients} and the middle panel of Fig.~\ref{fig:map_VVb}).

\subsection*{Population VI}

This group is made up of two main compact subgroups in the northern half of the region (Fig.~\ref{fig:map_VVa}). The eastern subgroup is labelled I in \citet{CantatGaudin19vela}, while the western subgroup is known as BH~23. Similarly to population V, this group exhibits a higher expansion rate in the direction of $\ell$ (or $X$) than in the vertical ($b$ or $Z$) direction.

\subsection*{Population VII}
This group is the second most populated after population IV. It is the densest, youngest, and most rapidly expanding group. Most of the stars identified by \citet{deZeeuw99} as members of the Vela~OB2 association belong to this group. 

Its spatial and kinematic substructure was studied in detail by \citet{CantatGaudin19vela}, who remarked that its location and dimension correlate with the expanding structure known as the IRAS Vela Shell \citep[IVS,][]{Sahu92phdt,Sahu93}. The authors suggested a common origin for these two objects in feedback from a supernova from an unidentified slightly older population. The newly discovered group labelled population III in this study is located inside the IVS and is $\sim$20\,Myr older than population VII.

\section{Discussion} \label{sec:discussion}
The main result of this study is that a large portion of the stars in the Vela-Puppis region are not concentrated in the few previously known clusters, but are rather distributed in sparse structures that are elongated along the Galactic plane.
They cluster in seven main groups in velocity space (top right panel of Fig.~\ref{fig:propermotionspace}). We note a correlation between the age and position of these groups, and their location in ($v_\ell$,$v_b$) space. The youngest group (population VII) is located near ($v_\ell$,$v_b$)$\simeq$(0\,km\,s$^{-1}$,-20\,km\,s$^{-1}$). The two groups with $v_\ell$>-20\,km\,s$^{-1}$ (populations V and VI) are located at low Galactic latitudes, while the others are located closer to the plane. This difference in velocity might be the result of a Galactic shear or the consequence of a velocity pattern that was already imprinted in the parent molecular cloud from which these young populations formed. We also note that the velocity $v_b$ seems to correlate with age, with the youngest group (population VII) having $v_b\sim0$\,km\,s$^{-1}$, while the oldest groups (populations II and IV) reach $v_b\sim-12$\,km\,s$^{-1}$. We show that Collinder~135, Collinder~140, and the recently discovered UBC~7 are all part of population IV and are therefore related to the Vela complex, although they lie outside of the area studied by \citet{deZeeuw99}, who defined the historical boundaries of the region that is referred to as the Vela~OB2 association.

All seven groups show clear signs of (mostly non-isotropic) expansion; the youngest group expands most rapidly. A mechanism that is commonly invoked to explain the significant expansion rates of young clusters is residual gas expulsion \citep[e.g.][]{Hills80,Lada84,Baumgardt07}, where the gas that has not turned into stars (but still contributes to the total potential well of the stellar cluster) is dispersed by the feedback of the most massive young stars. This mechanism is expected to produce isotropic expansion profiles, however, and therefore cannot be the only mechanism responsible for the observed velocity structure of the populations characterised in this study. 

Although we have shown beyond doubt that the overall spatial distribution of all identified populations is expanding, the \textit{Gaia}~DR2 we used is not sufficient to determine whether the denser clumps (the open clusters labelled in Fig.~\ref{fig:zari} as well as the dense subgroups of the youngest population) are bound, or are themselves slowly dispersing. Combining high-resolution spectroscopy and \textit{Gaia}-DR1 astrometry, \citet{Bravi18} have suggested that NGC~2547 might be in a supervirial state. A better characterisation could be obtained with \textit{Gaia}~DR2 data with a similar approach \citep[see in e.g.][for Cep~OB3]{Karnath18}. Interestingly, the observation that the distribution of young stars is more elongated along the $\ell$ direction than $b$ has been made by \citet{Eggen80}, who referred to the whole region as the ``Vela sheet''. The boundaries of the volume we investigated here were chosen to exclude the extended spur of young stars from the Scorpius-Centaurus complex (see Fig.~\ref{fig:zari}). We did not explore the possibility of a spatial or kinematical continuity between these two structures, although we note that the maps of \citet{Zari18} report a lack of young stars that would form a bridge from Scorpius-Centaurus to Vela.

The portrait of the region drawn by this study is that of a highly substructured ensemble of young stars, with seven main kinematic groups whose formation is spread over $\sim$35\,Myr. All of these groups physically extend over scales of several hundreds of parsecs. In agreement with \citet{Zari18} \citep[and at odds with][]{Bouy15}, we find no hint of a spatial gradient of age. This mixture of younger and older stars is indicative of multiple episodes of stellar formation over a short timescale. 
The chaotic aspect of the spatial distribution might reflect the turbulent structure of the primordial molecular cloud it formed from \citep[e.g.][]{Bonnell03,Girichidis11,Krumholz12}. The scale of the region as well as the velocity structure resemble the structures observed in numerical simulations of turbulent molecular clouds \citep[e.g.][]{Fujii15,Padoan16,Padoan17}. The age difference between the different populations is also similar to what is observed in supernovae-driven turbulent clouds (Padoan et al. in prep.). In this paradigm, the last episode of stellar formation (which formed population VII) would have been triggered by supernovae from the slightly older population II, which has the correct age and location for this scenario.

The volume of space investigated in this study is comparable with the dimensions of the nearby well-studied Scorpius-Centaurus (Sco-Cen) association. That region is composed of three main groups \citep[Upper Scorpius, Upper Centaurus Lupus, and Lower Centaurus Crux][]{Blaauw46,deZeeuw99,Wright18,Damiani18} with ages from $\sim$5 to $\sim$20\,Myr \citep[e.g.][]{Pecaut16}. It includes the two dense star-forming regions $\rho$~Ophiucus and Lupus (hosting stars younger than $\sim$1\,Myr). In order to explain that these two young groups are surrounded by older populations, \citet{Krause18} recently proposed a `surround and squash' scenario in which the dispersed population forms first in a dense region of a large elongated molecular cloud. The feedback from its massive stars generates a superbubble of hot pressurised gas that eventually expands and surrounds the residual gas. It finally compresses this residual gas into pockets of young stars that are surrounded by a sparser and older population. This scenario also naturally allows the young subgroups to exhibit different velocities.

The Vela-Puppis region we investigated here presents similarities with Sco-Cen in terms of dimensions. It also features distinct groups of different ages, and its youngest group (population VII) is located at the centre. Vela-Puppis appears to be much more substructured spatially, however. Another difference is that all seven populations we identified show clear signs of expansion and formed as dense clusters (even the older ones), while \citet{Wright18} have shown that the three main older groups of Sco-Cen are not expanding. Because Vela-Puppis is older than Sco-Cen, it is possible that the previous generation of stars that produced the superbubble that squashed the current populations into existence have now dispersed and can no longer be identified in velocity or positional space. Population I from this study could be the youngest representant of this hypothetical dispersed generation, and Vela-Puppis would be an older, more substructured, and scaled-up version of Sco-Cen.

An argument against the `surround and squash' scenario being at play in Vela-Puppis is that the various groups observed here are not organised in pockets that are isolated from each other, but are instead rather extended and overlap partially. That one group of stars (belonging to population II) inside the ring formed by the younger population VII, which \citet{CantatGaudin19vela} have shown to match the projected location of the IRAS Vela Shell, is also an argument in favour of a more standard triggered star formation mechanism such as supernovae explosions or a `collect and collapse' \citep[e.g.][]{Elmegreen77,Zavagno06,Dale07,Elmegreen11} origin for population VII.

\section{Conclusion} \label{sec:conclusion}

We used \textit{Gaia}~DR2 data to identify a large number of aggregates of young stars in a $\sim$600 square degree field in the direction of the Vela-Puppis constellations in the distance range 250 - 550\,pc. We grouped them into seven main kinematic populations. From comparisons with theoretical isochrones, the apparent ages of these groups range from $\sim$8 to 50\,Myr. The spatial structure of the groups extend across hundreds of parsecs, most of them with significant spatial and kinematic substructure. They all show clear signs of expansion along the Galactic $X$ and $Z$ axes (as determined from 3D velocities and from corrected tangential velocities), although this expansion is not isotropic for all groups, indicating that more than just internal mechanisms are at work. We observed no strong correlation between spatial location and age, but found correlations between kinematics (in particular velocity $v_b$) and age, suggesting that star formation took place in a turbulent environment rather than following a monotonic spatial sequence.

The limited number of radial velocities in the \textit{Gaia}-DR2 data as well as their relatively larger uncertainties prevent us from characterising the internal kinematics of the sub-components of the identified populations in greater detail. We therefore cannot determine whether the denser aggregates (the open clusters known prior to this study) are expanding or will remain gravitationally bound and remain clusters. Ground-based follow-up with high-resolution spectroscopy could provide precise radial velocity for a larger sample of stars and would also better constrain the stellar ages. This would help establish the time line for the stellar formation history of this complex region.

\section*{Acknowledgements}
We thank the anonymous referee for the helpful comments and constructive remarks that helped us to improve this manuscript.

TCG thanks E. Zari for providing the map adapted in Fig.~\ref{fig:zari}.

TCG acknowledges support from Juan de la Cierva - formaci\'on 2015 grant, MINECO (FEDER/UE). This work was supported by the MINECO (Spanish Ministry of Economy) through grant ESP2016-80079-C2-1-R (MINECO/FEDER, UE) and MDM-2014-0369 of ICCUB (Unidad de Excelencia 'María de Maeztu'). NJW acknowledges an STFC Ernest Rutherford Fellowship (grant number ST/M005569/1). PP acknowledges support  by the Spanish MINECO under project AYA2017-88754-P. VMP acknowledges support  by the Spanish MINECO under projects MDM-2014-0369 and AYA2017-88754-P.

The preparation of this work has made extensive use of Topcat \citep{Taylor05}, and of NASA's Astrophysics Data System Bibliographic Services, as well as the open-source Python packages Astropy \citep{Astropy13}, numpy \citep{VanDerWalt11}, and scikit-learn \citep{scikit-learn}.
The figures in this paper were produced with Matplotlib \citep{Hunter07}. This research has made use of Aladin sky atlas developed at CDS, Strasbourg Observatory, France \citep{Bonnarel00aladin,Boch14aladin}. 

This work presents results from the European Space Agency (ESA) space mission Gaia. Gaia data are being processed by the Gaia Data Processing and Analysis Consortium (DPAC). Funding for the DPAC is provided by national institutions, in particular the institutions participating in the Gaia MultiLateral Agreement (MLA). The Gaia mission website is \url{https://www.cosmos.esa.int/gaia}. The Gaia archive website is \url{https://archives.esac.esa.int/gaia}.

\bibliographystyle{aa} 
\linespread{1.5}               
\bibliography{biblio}

\clearpage
\appendix

\section{Maps} \label{appendix:maps}

\vfill
\begin{figure}[ht!]
\begin{center} \resizebox{\hsize}{!}{\includegraphics[scale=0.65]{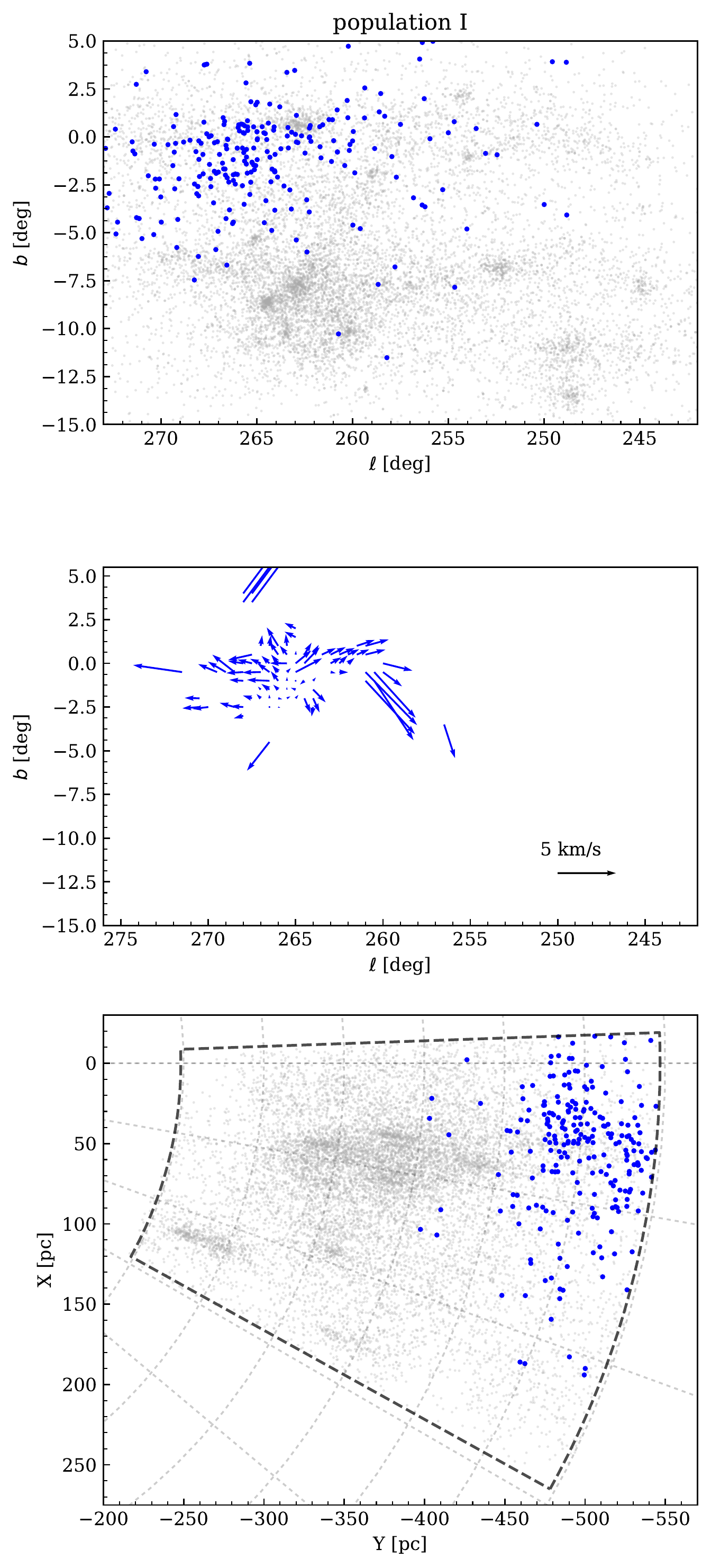}} \caption{\label{fig:map_Na} Top: Location of the population I stars on the sky. Middle: Vectors represent the local mean tangential velocity in a 0.5$^{\circ}$ radius (with respect to the group median) if at least three stars are present. The velocities are corrected for virtual contraction through the mean radial velocity of the population. Bottom: Location of the same stars projected on the Galactic plane. } \end{center}
\end{figure}

\vfill
\begin{figure}[ht!]
\begin{center} \resizebox{\hsize}{!}{\includegraphics[scale=0.65]{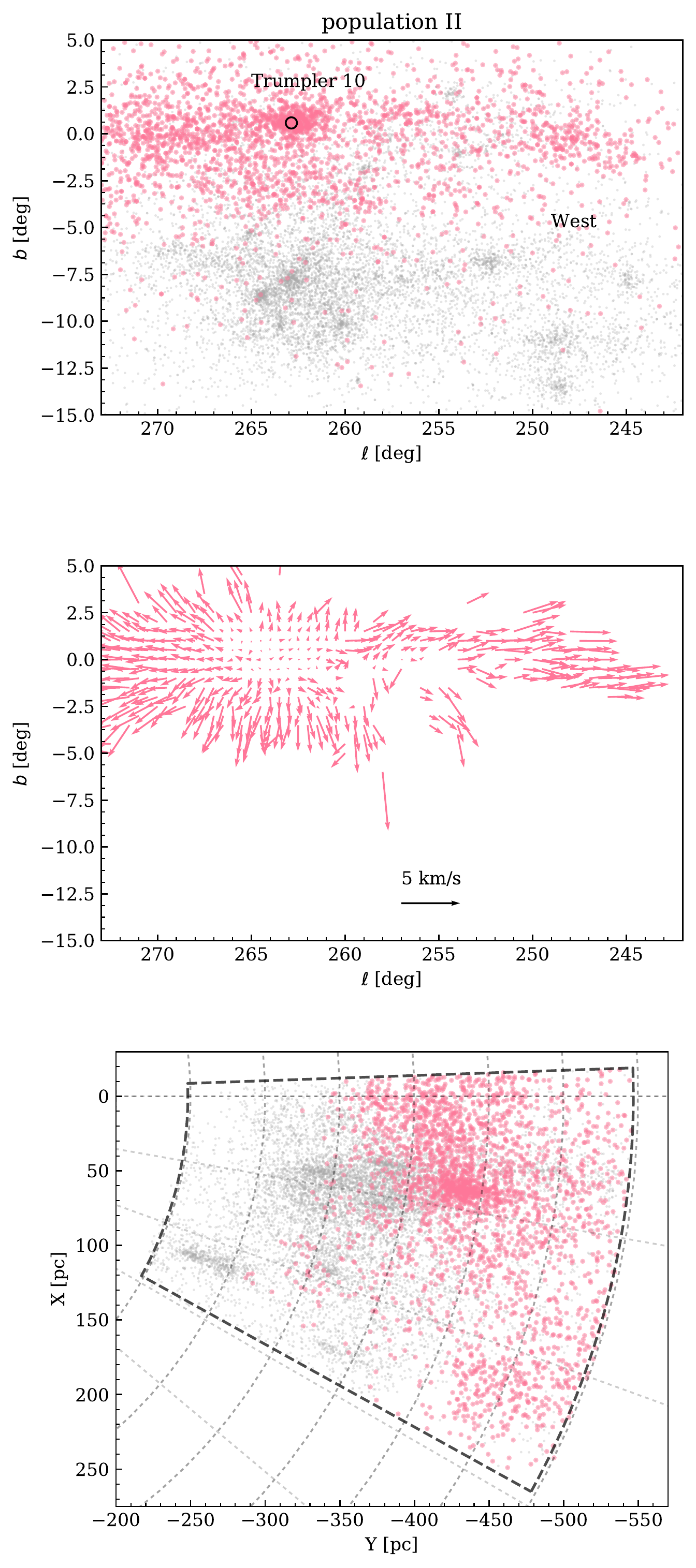}} \caption{\label{fig:map_Oa} Same as Fig.~\ref{fig:map_Na} for population II. } \end{center}
\end{figure}

\begin{figure}[ht]
\begin{center} \resizebox{\hsize}{!}{\includegraphics[scale=0.65]{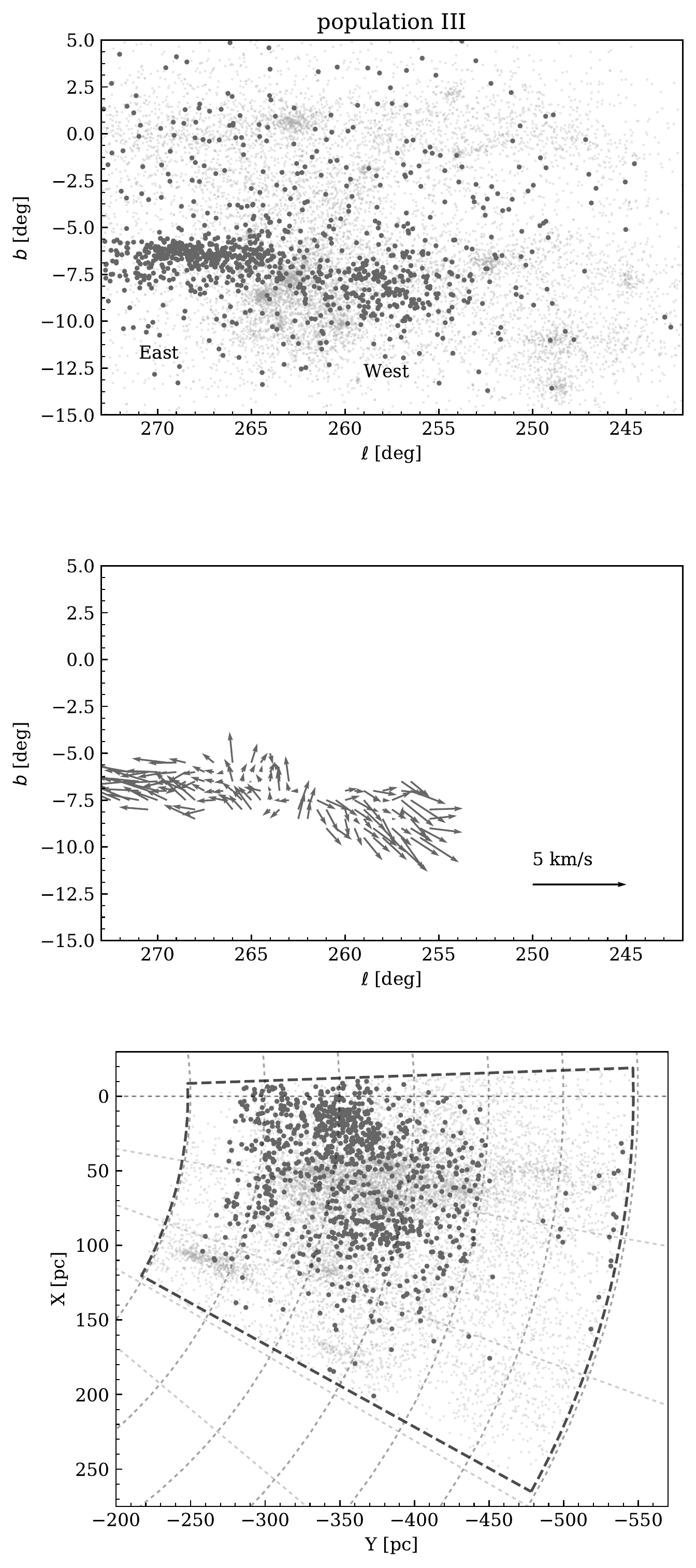}} \caption{\label{fig:map_Nb} Same as Fig.~\ref{fig:map_Na} for population III. } \end{center}
\end{figure}
\begin{figure}[ht]
\begin{center} \resizebox{\hsize}{!}{\includegraphics[scale=0.65]{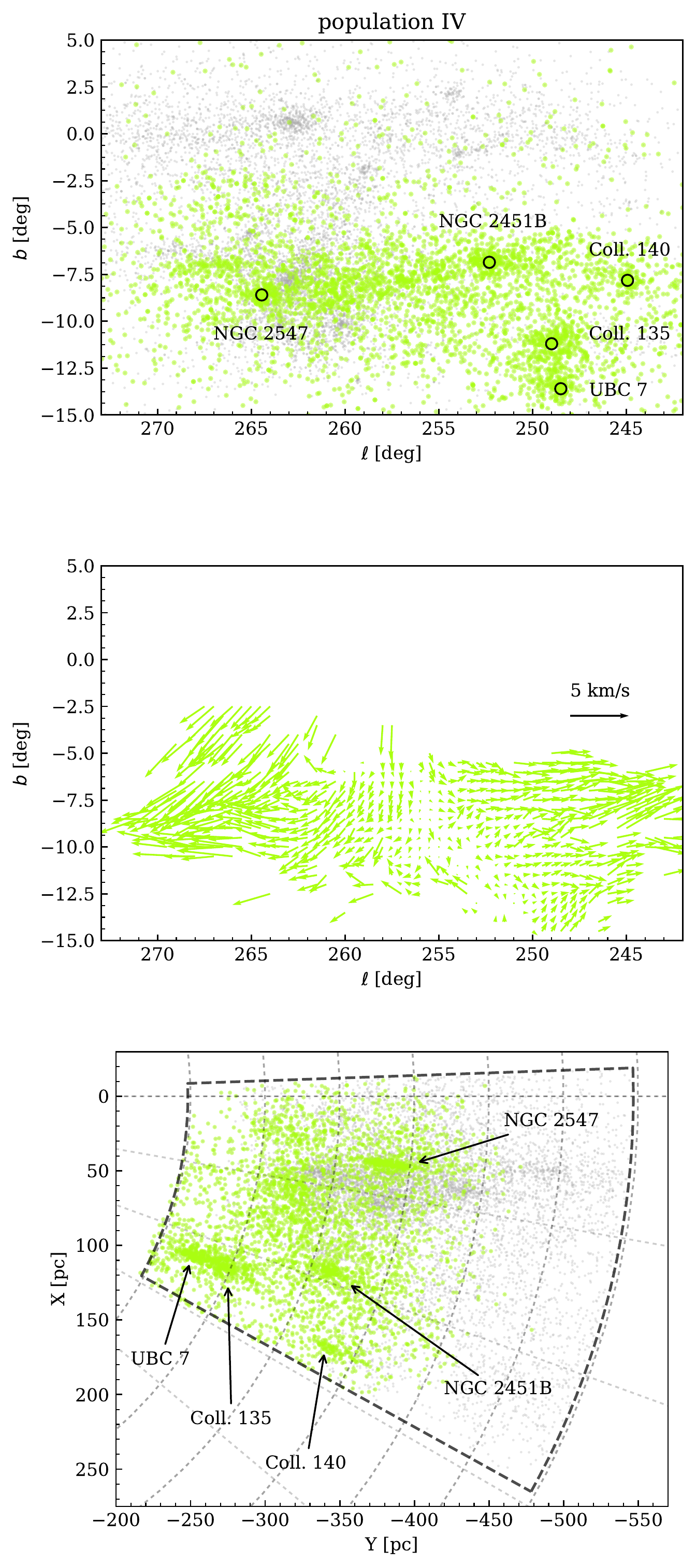}} \caption{\label{fig:map_Ob} Same as Fig.~\ref{fig:map_Na} for population IV. } \end{center}
\end{figure}
\begin{figure}[ht]
\begin{center} \resizebox{\hsize}{!}{\includegraphics[scale=0.65]{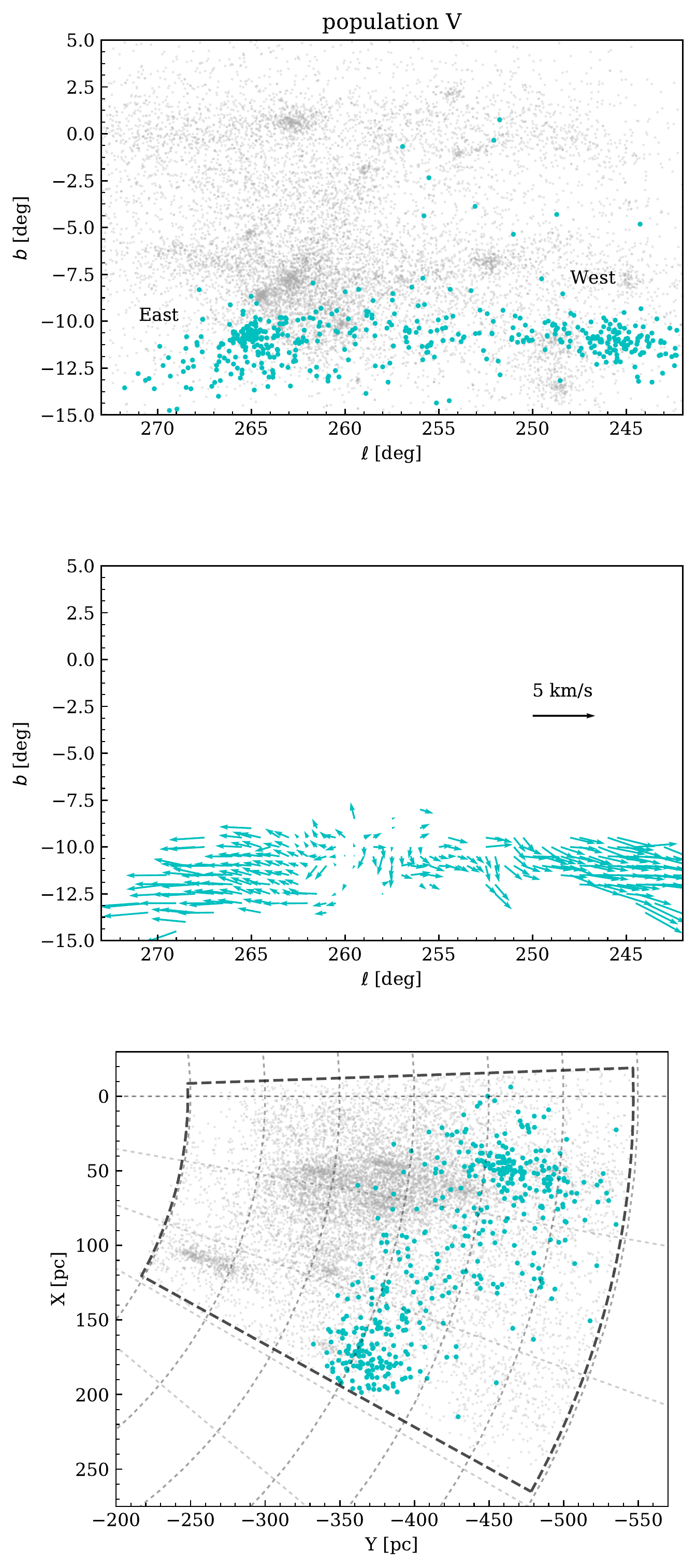}} \caption{\label{fig:map_VVb} Same as Fig.~\ref{fig:map_Na} for population V. } \end{center}
\end{figure}
\begin{figure}[ht]
\begin{center} \resizebox{\hsize}{!}{\includegraphics[scale=0.65]{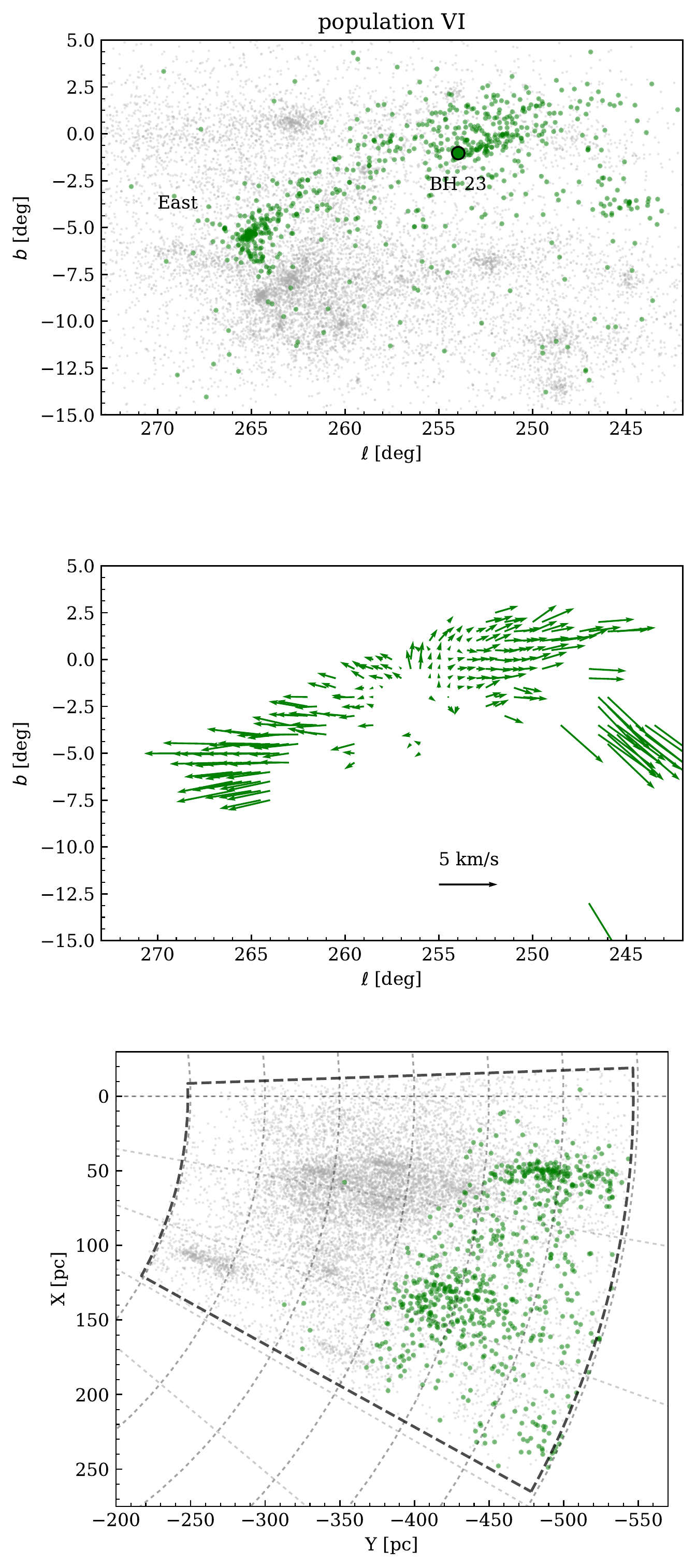}} \caption{\label{fig:map_VVa} Same as Fig.~\ref{fig:map_Na} for population VI. } \end{center}
\end{figure}
\begin{figure}[ht]
\begin{center} \resizebox{\hsize}{!}{\includegraphics[scale=0.65]{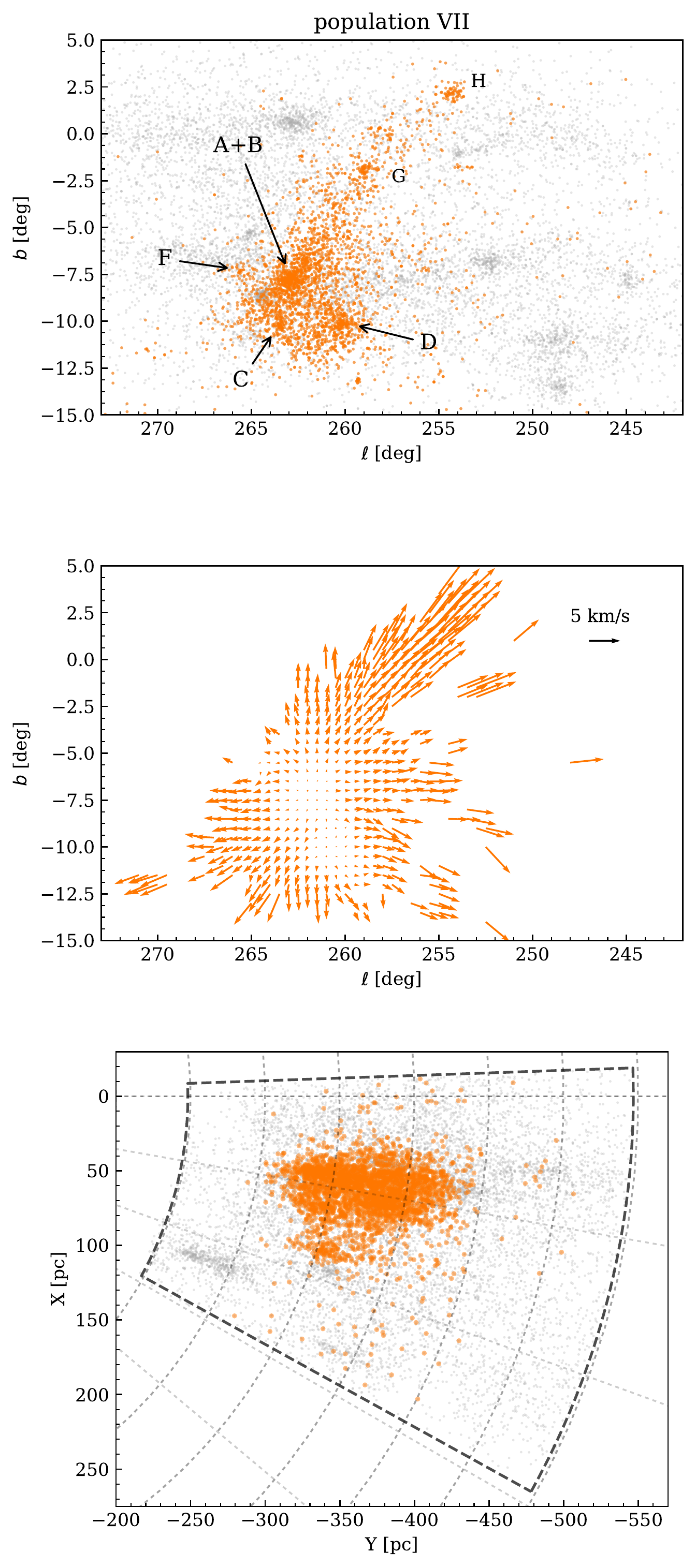}} \caption{\label{fig:map_V} Same as Fig.~\ref{fig:map_Na} for population VII. The subgroups are labelled as in \citet{CantatGaudin19vela}.} \end{center}
\end{figure}

\clearpage
\section{Extended component of population IV} \label{appendix:extended_popIV}
\begin{figure*}[ht]
\begin{center} \resizebox{\hsize}{!}{\includegraphics[scale=0.65]{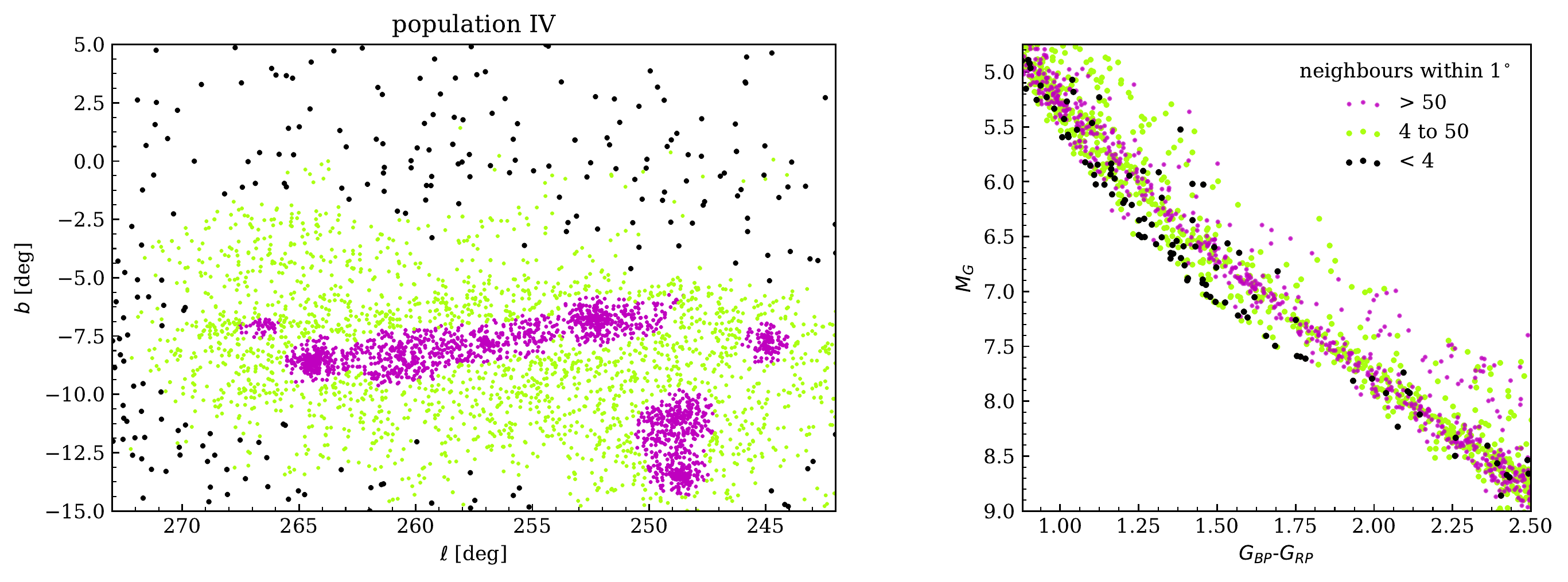}} \caption{\label{fig:map_extended_dense_cmd} Left: Members of population IV (same as top panel of Fig~\ref{fig:map_Ob}), colour-coded by local density (number of neighbours within one degree). Right: Colour-absolute magnitude diagram using the same colour code.} \end{center}
\end{figure*}

The extended distribution of stars around the known open clusters (NGC~2547, NGC~2451B, Collinder~135, UBC~7, and Collinder~140) of population IV might be interpreted at first glance as field contamination. In Fig.~\ref{fig:map_extended_dense_cmd} we compare the photometry (in a colour-absolute magnitude diagram) of stars in three density groups: the high-density clusters, the surrounding extended population, and the low-density outskirts.

The extended intermediate-density group perfectly overlaps with the known clusters, suggesting that these stars have the same age and are related to the clusters. The stars in the low-density outskirts, however, exhibit discrepant photometry and can therefore be considered as unrelated field stars that coincidentally share the same velocity as population IV stars. An a priori photometric selection of members is more difficult for population IV than for populations VI and VII, however, because of its older age (and therefore the proximity of the main-sequence field stars to the PMS stars in photometric space).

\clearpage

\section{Correcting for virtual contraction} \label{appendix:vtan}
\begin{figure}[ht]
\begin{center} \resizebox{\hsize}{!}{\includegraphics[scale=0.65]{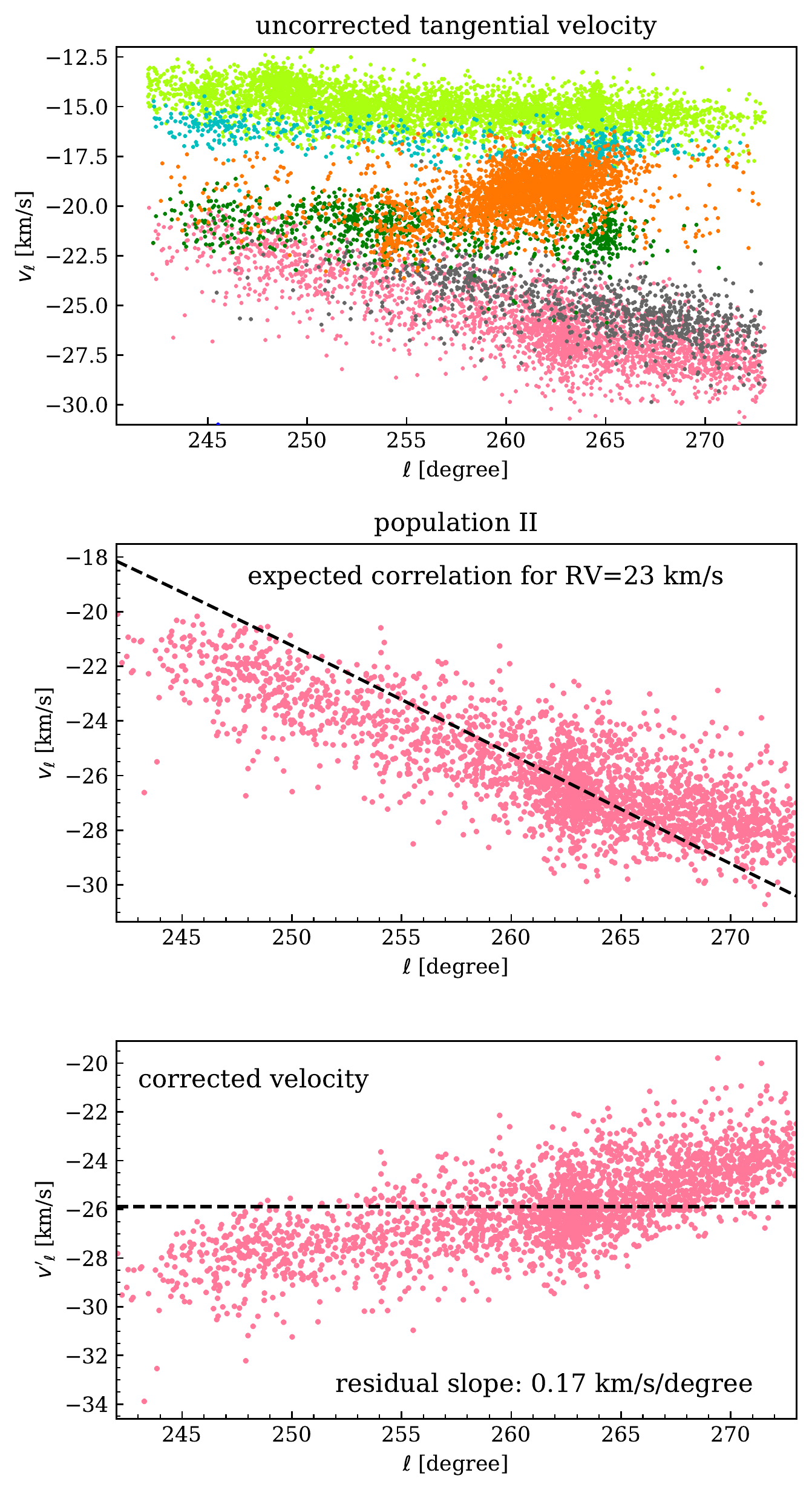}} \caption{\label{fig:virtual_Oa} Top: Observed tangential velocity $v_{\ell}$ as a function of Galactic longitude for populations II to VII. Middle: Same as top panel, showing only population II. The dashed line is the expected relation for a non-expanding group of stars with systemic velocity 23\,km\,s$^{-1}$ (shifted vertically to overlap with the observed data). Bottom: Corrected velocity (residuals of the observed velocities minus virtual contraction correction). } \end{center}
\end{figure}

The bulk motion of a group of stars moving away from us introduces a perspective effect that causes an apparent contraction of their spatial distribution, where the proper motion or tangential velocity vectors (when plotted with respect to the mean of the group) point inwards. This also results in a negative gradient of velocity $v_{\ell}$ with respect to Galactic longitude $\ell$ (and similarly, $v_{b}$ with respect to $b$). The top panel of Fig.~\ref{fig:virtual_Oa} shows that such apparent velocity gradients are visible in most of the identified populations.

The expected virtual motion of each star can be calculated as a function of the assumed bulk velocity of the group and the position of the star from the centre of the group. The equations we used in this study are those of \citet{Brown97} and are recalled in Sect.~\ref{sec:vtan}. 

In Fig.~\ref{fig:virtual_Oa} we show that for population II, the observed tangential velocity trend with $\ell$ is not as steep as expected from its mean radial velocity. In order to match the observed slope, we need to assume a radial velocity of 13\,km\,s$^{-1}$, which is largely beyond the uncertainty range (23$\pm$0.4\,km\,s$^{-1}$). The positive gradient in the residuals (bottom panel of Fig.~\ref{fig:virtual_Oa}) is indicative of expansion along the Galactic plane. 

We list in Table~\ref{table:slopes} the residual slopes observed after correction in the seven identified populations. The middle panels of Figs.~\ref{fig:map_Na} to \ref{fig:map_V} show that all of them exhibit significant expansion. The expansion rate of population VII is the highest and is stronger than its virtual contraction \citep{CantatGaudin19vela}.

\clearpage
\section{Extinction map} \label{appendix:ag}
\begin{figure}[ht]
\begin{center} \resizebox{\hsize}{!}{\includegraphics[scale=0.65]{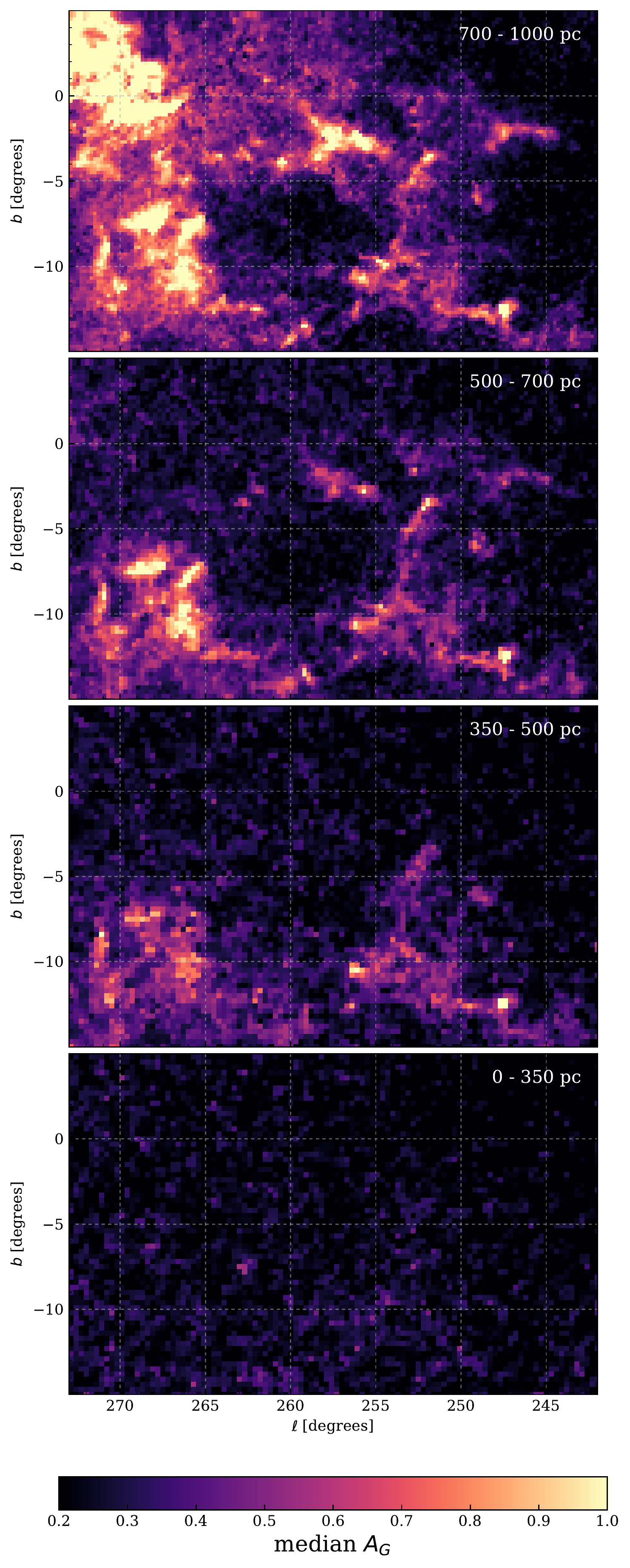}} \caption{\label{fig:ag_map} Median value of extinction ($A_G$) from the \textit{Gaia}~DR2 catalogue for stars in four different distance ranges. } \end{center}
\end{figure}

In this section we use the values of extinction $A_G$ provided in the \textit{Gaia}~DR2 catalogue \citep{Andrae18} to map the extinction in difference distances ranges, from 0 to 1\,kpc. To perform this experiment, we used all stars available in this volume of space and employed no photometric selection.
The result is shown in Fig.~\ref{fig:ag_map}. Although the individual values are noisy (with large uncertainties for individual stars) and overestimated for $A_G$ lower than $\sim$0.5 \citep{Andrae18,Sanders18}, these data are sufficient to show that stars within 350\,pc are virtually unaffected by interstellar extinction, and that stars with $b>-5^{\circ}$ are unaffected out to $\sim500$\,pc.

The few pixels of relatively higher extinction in near ($\ell$,$b$)=(263,-7.5) in the bottom panel of Fig.~\ref{fig:ag_map} correspond to the Gamma Velorum cluster, whose young PMS stars are interpreted as reddened main-sequence stars by the procedure of \citet{Andrae18}.

\end{document}